\documentclass[11pt]{article}
\usepackage{amsmath,amsthm,amssymb}
\usepackage{graphicx} 
\usepackage{amsmath,amssymb,latexsym}
\usepackage{psfrag,float,listings}
\usepackage{fancyvrb}
\usepackage{epsfig}
\usepackage{titlesec}
\usepackage{longtable}
\usepackage{rotating}
\usepackage{multirow}
\usepackage{multicol}
\usepackage{listings}
\usepackage{algorithm}
\usepackage{algpseudocode}
\usepackage{natbib}
\usepackage{bm}
\usepackage[caption = true]{subfig}
\usepackage{xcolor}
\usepackage{natbib}

\newcommand{\E}{\mathbb{E}}

\newcommand{\pr}{\mathrm{Pr}}

\newenvironment{theorem}[2][Theorem]{\begin{trivlist}
\item[\hskip \labelsep {\bfseries #1}\hskip \labelsep {\bfseries #2.}]}{\end{trivlist}}
\newenvironment{lemma}[2][Lemma]{\begin{trivlist}
\item[\hskip \labelsep {\bfseries #1}\hskip \labelsep {\bfseries #2.}]}{\end{trivlist}}

\newenvironment{definition}[2][Definition]{\begin{trivlist}
\item[\hskip \labelsep {\bfseries #1}\hskip \labelsep {\bfseries #2.}]}{\end{trivlist}}

\newcommand{\ignore}[1]{{}}
\newcommand{\blind}{1}

\begin{document}
 

\if1\blind
{
\title{\vspace*{-2cm}Network estimation via graphon with node features}
\author{
Yi Su\thanks{PhD candidate, Department of Statistics, University of California at Davis, Davis, CA 95616 (email: {\ttfamily njusu@ucdavis.edu}).}
\and Raymond K. W. Wong\thanks{Assistant Professor, Department of Statistics, Texas A\&M University, College Station, TX 77843 (email: {\ttfamily raywong@tamu.edu}).}
\and Thomas C. M. Lee\thanks{Professor, Department of Statistics, University of California at Davis, Davis, CA 95616 (email: {\ttfamily tcmlee@ucdavis.edu}).}
}

\date{\today}

\maketitle

\vspace*{-1cm}
\begin{abstract}
	Estimating the probabilities of linkages in a network has gained increasing interest in recent years. One popular model for network analysis is the exchangeable graph model (ExGM) characterized by a two-dimensional function known as a \textit{graphon}. Estimating an underlying graphon becomes the key of such analysis. Several nonparametric estimation methods have been proposed, and some are provably consistent. However, if certain useful features of the nodes (e.g., age and schools in social network context) are available, none of these methods was designed to incorporate this source of information to help with the estimation. This paper develops a consistent graphon estimation method that integrates the information from both the adjacency matrix itself and node features. We show that properly leveraging the features can improve the estimation. A cross-validation method is proposed to automatically select the tuning parameter of the method.
\end{abstract}


\section{Introduction}

A network (undirected simple graph) can be modeled as a partial observation of
an infinite random graph.
Exchangeable random graph model (ExGM) is a popular nonparametric model for infinite graphs where node indices are exchangeable
\citep[e.g.,][]{hoff2008modeling, kallenberg2006probabilistic, lovasz2012large,
	orbanz2015bayesian},
    i.e., the joint distribution of edges is invariant under permutation of node indices.
    For instance, in a social network when a node represents a person,
the assignment of node to person does not carry any information, and swapping the
node indices between any two people (i.e., relabeling) defines the same network.
An ExGM is characterized by a symmetric measurable function $w$ known as \textit{graphon} \citep{aldous1981representations,hoover1979relations},
which therefore plays a central role in model-based inference and prediction of network data under ExGM.
Based on the Aldous-Hoover Theorem, we assume the following generative model
of the network via graphon:
a set of latent labels $\{u_i\}$, each associated with a node, are first drawn independently from $\text{Uniform} (0,1)$.
These labels govern the probability of observing an edge between the corresponding two nodes through graphon. More specifically, given $u_i$ and $u_j$, the probability that there is a connection between the $i$-th node and the $j$-th node is given by $w(u_i,u_j)$.
According to these probabilities, edges will be then generated independently of each other conditional on $\{u_i\}$.

In general, graphon provides a unified and solid framework for modeling networks.  For instance, community structures widely used in the modeling of social networks correspond to a parametric piecewise-constant model of graphon.  More importantly, graphon opens up an opportunity for more flexible but challenging nonparametric modeling, which has sparked a recent surge of interest among researchers
\citep[e.g.,][]{airoldi2013stochastic,Wolfe-Olhede13, chan2014consistent,gao2015rate,zhang2015estimating, Klopp-Tsybakov-Verzelen17}, which is also the focus of the present work.  Since the knowledge of graphon facilitates our understanding of the underlying network, nonparametric graphon estimation helps discover unknown patterns in the corresponding network generation mechanism. Besides, statistical inference of network can also be conducted via graphon \citep[e.g.,][]{lloyd2012random, yang2014nonparametric}.

A typical assumption adopted by nonparametric graphon estimation is
the smoothness of the underlying graphon $w$.
If we were given the latent labels $\{u_i\}$,
the graphon estimation is simply a nonparametric regression problem.
However, $\{u_i\}$ are not observed, which poses a unique challenge.
Due to smoothness assumption, various researchers have made use of the idea that ``similar labels'' produce ``similar graphon slices'',
where a graphon slice at a label $u\in[0,1]$ is a one-dimensional function $w(u,\cdot)$.
In other words, if the labels $u_i$ and $u_j$ of two nodes are close, $w(u_i,
\cdot)$ and $w(u_j, \cdot)$ should be similar.
As graphon is unknown, several methods
\citep[e.g.,][]{airoldi2013stochastic, chan2014consistent, gao2015rate} instead use the rows and/or columns of the adjacency matrix as a proxy of graphon slices,
to describe the distance between nodes, based upon which smoothing procedures can be constructed.
Despite the success of many existing methods relying solely on the adjacency matrix,
very often features (or attributes) of the nodes are available aside from
the network (adjacency matrix) itself, and could potentially provide important information
for network estimation.
Take Facebook friendship network as an example. 
It is conceivable that users who share similar values
of certain features (e.g., age and schools) will have similar connection behaviors.
These additional features can be valuable resource
for better estimating the underlying probabilities of linkages and the graphon.

The main contribution of the paper is the proposal of a nonparametric graphon estimation method that is capable of utilizing the information hidden in the node features for better network estimation. This can be realized by relating node features to graphon slices through the latent  labels. That is, close labels should correspond to both similar graphon slices and similar node features.

The rest of this paper is organized as follows. In Section~\ref{sec:background}, we will review some basic definitions and related work on graphon estimation,
and summarize our contribution.
We introduce the proposed framework and estimation method in Section~\ref{sec:method}. Theoretical results are provided in Section~\ref{sec:theory},
while numerical experiments on synthetic graphons are shown in
Section~\ref{sec:experiment}.
Finally, we apply our method to a real-world friendship network in Section \ref{sec:application}, and supplementary material is deferred to the appendix.

\section{Background}
\label{sec:background}

This section presents necessary background material. In sequel, for any matrix $\bm{M}$, we use $M_{ij}$, $\bm{M}_{i\cdot}$ and $\bm{M}_{\cdot j}$ to denote its $(i,j)$-th element, $i$-th row and $j$-th column, respectively.

\subsection{Graphon, exchangeability, and identifiability}
Let $\bm{A}\in\{0,1\}^{n\times n}$ be the adjacency matrix of a non-directed simple graph with $n$ nodes ($n$ can be infinity); i.e., $A_{ij}=1$ if the $i$-th node and $j$-th node is connected, and 0 otherwise. For an infinitely sized graph, we say it is \textit{exchangeable} if the distribution over $\bm{A}$ is invariant under any permutation of nodes. The Aldous-Hoover theorem \citep{aldous1981representations, hoover1979relations} guarantees that every ExGM must be represented by a graphon.
\begin{definition}{2.1} (Graphon)
	A graphon is a symmetric measurable function $w: \left[0,1\right] ^2 \rightarrow\left[0,1\right]$ such that
	\begin{equation*}
	\pr(A_{ij}=1 \vert u_i,u_j)=w(u_i,u_j),
	\end{equation*}
	where $u_i \stackrel{iid}{\sim}\text{Uniform}(0,1)$ for $i\in\mathbb{N}$.
\end{definition}
A network of size $n$ can be modeled as a partial observation of an ExGM, and thereby generated by the following two-step sampling scheme:
\begin{equation}
\label{eqn:generate}
\aligned
u_i &\stackrel{iid}{\sim} \text{Uniform} (0,1), \quad  i=1,2,\ldots,n;\\
A_{ij} \vert u_i, u_j &\stackrel{ind}{\sim} \text{Bernoulli} (w(u_i,u_j)), \quad i<j.
\endaligned
\end{equation}

Identifiability is a well-known issue of graphon, and different graphons can give rise to the same ExGM. Specifically, up to a measure preserving transformation $\varphi$, $w'(u,v) := w(\varphi(u),\varphi(v))$ and $w$ define the same ExGM. That is, the distributions of these two random arrays are the same.
To guarantee an unique representation, one can impose
the \textit{strict monotonicity of degree condition} \citep{bickel2009nonparametric,yang2014nonparametric},
which assumes that
there exists a measure preserving transformation $\varphi$ such that $w^{can}(u,v):=w(\varphi(u),\varphi(v))$, and
\begin{equation}
g^{can}(u)=\int_{0}^{1}w^{can}(u,v)dv
\label{eqn:mono}
\end{equation}
is strictly increasing in $u$. Here $w^{can}$ is called the \textit{canonical} form of graphon $w$, and is a unique representation of the underlying ExGM.
However, this assumption is restrictive because it excludes commonly used models such as the stochastic block model. In our framework, we will not enforce strictly monotonic node degrees.

In principle, one cannot determine which graphon, from its equivalence class,
that generates the underlying network based on the adjacency matrix.
There are two layers of estimation: the first layer is the estimation of the
graphon $w$ while the second layer is the estimation of the latent labels
$\{u_i\}$. Since it is unrealistic to estimate the labels without strong
assumptions, the main purpose of estimating a graphon is sometimes to obtain the
probabilities of linkages at the observed nodes, $(w(u_i,u_j))_{1\le i,j\le n}$.
This is also the goal of this paper.

\subsection{Related work}
In the literature of graphon estimation, a commonly adopted strategy is to employ the graphon slices (which can be estimated by the rows and columns of the adjacency matrix) to define the distance between nodes. With this, one can group similar nodes together into different blocks and estimate the graphon values within any block by averaging the number of edges in it.

\cite{airoldi2013stochastic} proposed the Stochastic Blockmodel Approximation (SBA) algorithm, which approximates the graphon by a piecewise constant function.
Their estimator is consistent in mean squared error, but a key assumption is that there are at least $2T$ ($T\in\mathbb{N}_+$) independent realizations generated from $w$, which is unlikely to hold in reality. They group the nodes into $K$ blocks, and the estimated graphon is a piecewise constant function over $K\times K$ blocks. 

Some other methods are based on the strong assumption of strict monotonicity of
degree \citep{chan2014consistent, yang2014nonparametric},
under which a canonical graphon is well-defined and hence treated as the estimand of interest.
One representative of this category is the Sorting-and-Smoothing (SAS) algorithm
proposed by \citet{chan2014consistent}. It first sorts the nodes according to
their empirical degrees, then computes a local histogram estimator
$\hat{\bm{H}}\in[0,1]^{k\times k}$ for some bandwidth $h=n/k$, and finally applies a
smoothing technique to obtain the final estimate. This SAS estimator is
consistent and reaches the rate of convergence $n^{-1}\log n$.
This rate matches with the optimal rate
in general graphon estimation
without the assumption of strict monotonicity of degree
\citep{gao2015rate}.

Another popular method is
Universal Singular Value Thresholding (USVT) proposed by
\citet{chatterjee2015matrix},
which targets at general matrix denoising problems with missing values. Since
this method is not specifically for graphon, the rate of convergence is not competitive.

More recently, \citet{zhang2015estimating} proposed a novel Neighborhood Smoothing (NBS) method for estimating the underlying probability matrix $P_{ij}$, which is equivalent to estimating the graphon $w(u_i,u_j)$. Different from the SBA and the SAS algorithms, these authors proposed an adaptive neighborhood selection method which allows each node to have its own neighbors. The NBS method performs very well for a wide range of graphons in both low-rank and high-rank situations, and the only assumption on graphon is piecewise Lipschitz. These authors also showed that the error rate of NBS
is the smallest among all existing non-combinatorial methods.

\subsection{Our contribution}
To the best of our knowledge, none of the existing methods are designed to utilize
information other than the adjacency matrix itself
for nonparametric newtork/graphon estimation.
With additional node features, the estimation could be
much improved.
In this paper, we propose a generative model of node features
which allows borrowing information from the features
in an adaptive manner to improve the network/graphon estimation.
If similar node features correspond to similar graphon slices,
these features are valuable, especially when the graphon itself has weak/local signals.
In contrast, it could happen that two nodes with
identical attributes behave very differently.
In such scenarios, it is unwise to contaminate the estimation by using these
unhelpful features. We will avoid this contamination by selecting the tuning parameter adaptively, which
controls the weight of using the features' information.

\section{Methodology}
\label{sec:method}
We begin with some notations. Recall that $\bm{A}\in\{0,1\}^{n\times n}$ is an observed adjacency matrix generated by graphon $w(u,u)$. That is,
\begin{equation*}
A_{ij} \vert u_i, u_j \stackrel{ind.}{\sim} \text{Bernoulli} (w(u_i,u_j))
\end{equation*}
where $u_i \stackrel{iid}{\sim} \text{Uniform} (0,1)$ for $i=1,\ldots,n$. For each node $i$, we also observe a feature vector $\bm{X}_i\in\mathbb{R}^p$, $i=1,\ldots,n$. We assume that these $\bm{X}_i$'s are also generated from (unknown) latent labels:
\begin{equation}
\bm{X}_i = f(u_i) + \bm{e}_i,
\label{eqn:fpluse}
\end{equation}
where $f = (f_1, ..., f_p)^T:\mathbb{R}\rightarrow\mathbb{R}^p$ is an unknown function,
and $\bm{e}_i$ is a random vector with independent entries of mean zero and variance $\sigma^2$. 
Also, $\{\bm{e}_1,\dots, \bm{e}_n, u_1,\dots, u_n\}$ are mutually independent.
We note that $X_{i1}, \dots,X_{ip}$, the elements of $\bm{X}_i$ are dependent in general due to the sharing of $u_i$ through $f_1,\dots, f_p$;
and the assumption of constant variance $\sigma^2$ can be relaxed easily.
In addition, we assume that $w$ and $f$ are piecewise Lipschitz functions which will be defined in Section \ref{sec:theory}.

Although we aim to utilize the features for better graphon estimation,
our feature model (\ref{eqn:fpluse}) is fairly general and
does not assume the usefulness of features for graphon estimation in priori.
To see this, the essential information of $\bm{X}_i$ for graphon estimation
is captured by the hidden label $u_i$.
When $f$ is monotonic, close feature vectors correspond to close latent labels, which will generate similar slices in the adjacency matrix (under smoothness assumption of $w$). In this case, similarity of features is a helpful source we can borrow information from. On the other hand, when $f$ is non-monotonic, close features does not necessarily imply similar graphon slices. For example, if $f(u_1)=f(u_2)$ for very different $u_1$ and $u_2$, then whether including feature similarity is useful or not will depend on if graphon slice $w(u_1,\cdot)$ is close to $w(u_2,\cdot)$. Given that we do not know if the latter is true, the use of features could worsen the graphon estimation.
In the subsequent sections, we will develop a method that allows adaptive incorporation of feature information
via a tuning parameter, as well as a data-adaptive method for choosing such a parameter.

In what follows, we use $\bm{P}$ to denote the underlying (conditional) probability matrix with $P_{ij} = w(u_i, u_j)$, i.e., $\bm{P}=\E(\bm{A} | \{u_i\}_{i=1}^n)$.
Our goal is to estimate $\bm{P}$.

\subsection{Feature Assisted Neighborhood Smoothing (FANS)}
This subsection provides a general description of the proposed method for graphon estimation.  The method is called FANS, short for Feature Assisted Neighboring Smoothing.

Since the latent labels $\{u_i\}$ are unavailable, the key of estimating a graphon is to define a measure of node dissimilarity. Here we define a (squared) dissimilarity function $d(i,j)$ between the $i$-th node and the $j$-th node ($i\neq j$) as the weighted sum of two terms: 
\begin{equation}
\label{eqn:defined}
d^2(i,j) = d_0^2(i,j) + \lambda s^2(i,j),
\end{equation}
where their relative weights are determined by a tuning parameter $\lambda\ge 0$.
In (\ref{eqn:defined}), $d_0(i,j)$ is a distance measure for graphon slices
while $s(i,j)$ is a distance measure for features;
exact forms of these two measures are given in Section \ref{sec:d0s}. The
parameter $\lambda$ controls how much information we want to borrow from the
node features. When these features are not helpful, we could avoid using them
by setting $\lambda=0$. We will discuss a data-driven choice of $\lambda$ later.

The first step of the proposed estimation method is to estimate $d(i,j)$.  Once such an estimate $\hat{d}(i,j)$ is obtained, the next step is to obtain the neighborhood for each node. Similar to the NBS method \citep{zhang2015estimating}, we define the neighborhood of the $i$-th node as
\begin{equation}
N_i = \{i'\neq i: \hat{d}(i,i')\leq q_i(h) \},
\label{eqn:defineN}
\end{equation}
where $q_i(h)$ is the $h$-th sample quantile of the set $\{\hat{d}(i,i'): i'\neq i \}$, and $h=C_0\sqrt{{\log n}/{n}}$ with a global constant $C_0>0$.
From our experience, the performance of the proposed method is not sensitive to the choice of $C_0$ in a mild range between 0.5 and 1.5. In practice, we set $C_0=1$. Unlike the SBA and SAS algorithms, this neighborhood is different from node to node. Finally, the estimated graphon evaluated at $(u_i,u_j)$ is given by
\begin{equation}
\hat{w}(u_i,u_j) = \hat{P}_{ij} = \frac{1}{2}\left( \frac{\sum_{i'\in N_i} A_{i'j}}{|N_i|} + \frac{\sum_{j'\in N_j} A_{ij'}}{|N_j|} \right).
\label{eqn:defineestw}
\end{equation}
To sum up, the proposed FANS method consists of the following three major steps:
\begin{enumerate}
	\item Obtain an estimate for $d(i,j)$ in (\ref{eqn:defined}), where $d_0(i,j)$, $s(i,j)$ are estimate by (\ref{def:d0hat}) and (\ref{def:shat}), and $\lambda$ is chosen by Algorithm \ref{alg:cv}.
	\item Calculate the $N_i$ for all $i$ using (\ref{eqn:defineN}).
	\item Compute the estimated $w(u_i,u_j)$ with (\ref{eqn:defineestw}).
\end{enumerate}
Details for these three steps are given below.  See also Algorithm~\ref{alg:proposed}.

\subsection{Defining $d_0$ and $s$}
\label{sec:d0s}
Following the ideas of \cite{airoldi2013stochastic} and \cite{zhang2015estimating}, we define $d_0$ using the $L_2$ distance of graphon slices. To be more specific, for any $i\neq j$, define 
\begin{equation*}
\aligned
d_0^2(i,j) &= \int_{0}^{1}|w(u_i,v)-w(u_j,v)|^2 dv\\
&= \langle w(u_i,\cdot), w(u_i,\cdot)\rangle + \langle w(u_j,\cdot), w(u_j,\cdot)\rangle - 2\langle w(u_i,\cdot), w(u_j,\cdot)\rangle,
\endaligned
\end{equation*}
where
\begin{equation*}
\langle g_1(x), g_2(x)\rangle := \int_{0}^{1}g_1(x)g_2(x)dx.
\end{equation*}
With slight notational abuse, we use the notation $\langle\cdot,\cdot\rangle$ to denote both the $L_2$ inner product of two functions and the Euclidean inner product of two vectors.

We immediately notice that the last term $\langle w(u_i,\cdot), w(u_j,\cdot)\rangle$ can be estimated by $\langle \bm{A}_{i\cdot}, \bm{A}_{j\cdot} \rangle /n$,
because the entries of $\bm{A}_{i\cdot}$ and $\bm{A}_{j\cdot}$ are ``almost'' independent (except for $A_{ij}$ and $A_{ji}$). However, $\langle w(u_i,\cdot), w(u_i,\cdot)\rangle$ cannot be well estimated using $\langle \bm{A}_{i\cdot}, \bm{A}_{i\cdot} \rangle /n$ (one can consider estimating $p^2$ in Bernoulli distribution as an analogy.) Similarly for $\langle w(u_j,\cdot), w(u_j,\cdot)\rangle$. The SBA algorithm solves this issue by requiring that at least two independent copies of the network are observed.

Following \citet{zhang2015estimating}, we instead use an approximate upper bound of $d_0^2(i,j)$,
which is motivated by the following heuristic argument.
First,
\begin{equation}
d_0^2(i,j) = \langle w(u_i,\cdot)-w(u_j,\cdot), w(u_i,\cdot)\rangle - \langle w(u_i,\cdot)-w(u_j,\cdot), w(u_j,\cdot)\rangle.
\label{eqn:d00}
\end{equation}
With large sample, it is likely that
there exist $\tilde{i}$, $\tilde{j}$ such that
$|u_{\tilde{i}}-u_i|\le \varepsilon$ and $|u_{\tilde{j}}-u_j|\le \varepsilon$
for small $\varepsilon$.
Suppose $w(u,u')$, as a function of $u'$, has a Lipschitz constant $L$ for every $u\in[0,1]$.
For the first term of \eqref{eqn:d00}, we have
\begin{equation*}
\aligned
| \langle w(u_i,\cdot)-w(u_j,\cdot), w(u_i,\cdot)\rangle| &= |\langle w(u_i,\cdot)-w(u_j,\cdot), w(u_{\tilde{i}},\cdot)\rangle + \langle w(u_i,\cdot)-w(u_j,\cdot), w(u_i,\cdot)-w(u_{\tilde{i}},\cdot) \rangle |\\
&\leq \max\limits_{k\neq i,j} |\langle w(u_i,\cdot)-w(u_j,\cdot), w(u_k,\cdot) \rangle| + L \varepsilon,
\endaligned
\end{equation*}
since $\int^1_0 (w(u_i,u')-w(u_j,u'))^2 du'\le 1$
and $\int^1_0 (w(u_i,u')-w(u_{\tilde{i}},u'))^2 du'
\le L^2\varepsilon^2$.
Similarly for the second term of \eqref{eqn:d00}.
Therefore, we have
\begin{equation*}
d_0^2(i,j) \leq 2\max\limits_{k\neq i,j} |\langle w(u_i,\cdot)-w(u_j,\cdot), w(u_k,\cdot) \rangle| + 2L\varepsilon.
\end{equation*}
Disregarding the multiplicative constant and the small term $\epsilon$,
it can be estimated by
\begin{equation*}
\tilde{d}_0^2(i,j) := \max\limits_{k\neq i,j}|\langle \bm{A}_{i\cdot}-\bm{A}_{j\cdot}, \bm{A}_{k\cdot} \rangle| / n.
\end{equation*}
In the same vein, we define $s^2(i,j) = \|f(u_i) - f(u_j)\|^2$, and the estimator of its upper bound (up to a multiplicative constant) is
\begin{equation}
\label{def:shat}
\hat{s}^2(i,j) := \max\limits_{k\neq i,j}|\langle \bm{X}_{i}-\bm{X}_{j}, \bm{X}_{k} \rangle| / p,
\end{equation}
where $\bm{X}_i\in\mathbb{R}^p$ is the feature vector for the $i$-th node. The usage
of these upper bounds and their estimates will be justified both theoretically
and empirically in subsequent sections.

\subsection{Tie-corrected $\tilde{d_0}$}
Due to the nature of upper bound and the fact that $A_{ij}$'s are binary, $\tilde{d}_0(i,j)$ has an issue of ties. For now suppose $\lambda=0$; i.e., not using any node feature. When $w$ is a piecewise constant function (for which the stochastic block model is an example), $D_i := \{ \tilde{d}_0(i,j):j\neq i \}$ may contain a large number of ties. In our simulation we found that when $n=500$, the number of unique values in $D_i$ can be as small as 65. These ties cause a problem when defining $N_i$ because the set of boundary points $B_i := \{i'\neq i: \tilde{d}_0(i,i')=q_i(h) \}$ can be large. Hence $N_i$ does not change continuously as $h$ changes, and therefore it may include too many nodes on the boundary. Clearly, not all the nodes in $B_i$ are as useful as those in $D_i \backslash B_i$, but there is no mechanism to distinguish which nodes in $B_i$ are useful to be included. On the other hand, we do not want to either exclude or include all of them. To solve this problem, we propose using an adjusted $\tilde{d_0}$ by applying a random perturbation to the original definition:
\begin{equation}
\label{def:d0hat}
\hat{d}_0^2(i,j) := \max\limits_{k\neq i,j} \left( |\langle \bm{A}_{i\cdot}-\bm{A}_{j\cdot}, \bm{A}_{k\cdot} \rangle| + (t/n) \right) / n,
\end{equation}
with $t\sim\text{Uniform}(0,1)$. Since $|\langle \bm{A}_{i\cdot}-\bm{A}_{j\cdot}, \bm{A}_{k\cdot} \rangle|$ is always an integer, this randomization will not change the order of any other points that are not ties.
As for $\lambda>0$, the randomization is not necessary if $\bm{X}_i$'s are continuous random variables
since ties are unlikely.

\begin{algorithm}[ht]
	\caption{The FANS method}
	\label{alg:proposed}
	\begin{algorithmic}[0]
		\State \textbf{Input}: $\bm{A}\in\{0,1\}^{n\times n}$, $\bm{X}=(\bm{X}_1,\ldots,\bm{X}_{n})^T\in\mathbb{R}^{n\times p}$, $\lambda$
		\State \textbf{Output}: $\hat{\bm{P}}$
		\State Step 1: Calculate $\hat{d}_0^2(i,j) = \max\limits_{k\neq i,j} \left( |\langle \bm{A}_{i\cdot}-\bm{A}_{j\cdot}, \bm{A}_{k\cdot} \rangle| + t/n \right) / n$ with $t\sim\text{Uniform}(0,1)$.
        \State Step 2: Calcualte $\hat{s}^2(i,j) = \max\limits_{k\neq i,j}|\langle \bm{X}_{i}-\bm{X}_{j}, \bm{X}_{k} \rangle| / p$.
        \State Step 3: Compute $\hat{d}^2(i,j) = \hat{d}_0^{2}(i,j) + \lambda \hat{s}^2(i,j)$.
		\State Step 4: Define $N_i=\{ i'\neq i: \hat{d}(i,i') < q_i(h) \}$ where $h=C_0\sqrt{\frac{\log n}{n}}$.
		\State Step 5: Output $\hat{P}_{ij} = \frac{1}{2}\left( \frac{\sum_{i'\in N_i} A_{i'j}}{|N_i|} + \frac{\sum_{j'\in N_j} A_{ij'}}{|N_j|} \right)$.	
	\end{algorithmic}
\end{algorithm}

\subsection{Cross-validation for selecting $\lambda$}
\label{sec:cv}
Selection of any tuning parameter in network estimation is generally a
challenging problem. Popular data-spliting strategies like cross-validation has no trivial extension to the setting of network data.
Recently, \citet{chen2016network} proposed a piecewise node-pair splitting
technique for cross-validation to determine the number of communities $K$
in a stochastic block model. An unpublished work of \citet{li2016network}
proposed a two-stage network cross-validation by edge splitting for stochastic block model. A key assumption of both methods is that $P$ is low-rank, which does not fit in general graphon framework.

One advantage of having node features is that, by solely comparing the features,
it is possible to generate prediction of edge connection.
If a new node $i$ comes into an
existing network, we can find its nearest neighbor $i^*$ based on its features.
Then we use the estimated graphon slice of $i^*$ as a prediction of $i$'s
connections. Assuming that the features are useful (which means the optimal
$\lambda\neq 0$), a good model should predict $i^*$ with a small error.

Our cross-validation method is outlined in Algorithm \ref{alg:cv}. It would be natural to use
$\ell_2$ norm or negative binomial log-likelihood as the loss function.
However, we found that $\ell_2$ norm would fail easily since it is much less
robust than $\ell_1$ error. The reason we do not use log-likelihood is that we
may have $\hat{P}_{ij}=0$ or 1 occasionally. Simulation results suggest that
our method works well and will set $\lambda\approx 0$ in cases where node
features are not helpful (such as Graphon 2 in Section \ref{sec:experiment}.)

\begin{algorithm}[ht]
	\caption{Cross-validation for choosing $\lambda$}
	\begin{algorithmic}[0]
		\State \textbf{Input}: $\bm{A}\in\{0,1\}^{n\times n}$, $\bm{X}=(\bm{X}_1,\ldots,\bm{X}_{n})^T\in\mathbb{R}^{n\times p}$, $\Lambda = \{\lambda_1,\ldots,\lambda_Q\}$.
		\State \textbf{Output}: Optimal $\lambda_{opt}$.
		\For{$m=1,\ldots,M$}
		\State Randomly sample $[10\%n]$ nodes $V \subset \{1,\ldots,n\}$ as the validation set.
		\State Let $T=\{1,\ldots,n\} \backslash V$ be the training set.
		\State Split $\bm{A}$ and $\bm{X}$: $\bm{A} = \begin{pmatrix}
		\bm{A}^{(T)} & \bm{A}^{(T\times V)}\\
		\bm{A}^{(V\times T)} & \bm{A}^{(V)}
		\end{pmatrix}$, 
		$\bm{X} = \begin{pmatrix}
		\bm{X}^{(T)} \\ \bm{X}^{(V)}
		\end{pmatrix}$.
		\For{$q = 1,\ldots, Q$}
		
		\State Fit $\hat{P}_{ij}$ for $(i,j)\in T\times T$ using $\bm{A}^{(T)}$, $\bm{X}^{(T)}$ and $\lambda_q$.
		\For{$i\in V$}
		\State Find $i^* = \arg\min\limits_{i'\in T}\Vert \bm{X}_{i'} - \bm{X}_i\Vert_2$, and let $\hat{P}_{ij} = \hat{P}_{i^*j}$, $j\in T$.
		\EndFor
		
		\State Compute loss for model $q$, $L_q^{(m)} = \frac{1}{|V||T|}\sum\limits_{(i,j)\in V\times T} | A_{ij}-\hat{P}_{ij} |$.
		\EndFor
		
		\EndFor
		
		\State Let $L_q = \frac{1}{M}\sum\limits_{m=1}^{M}L_q^{(m)}$, and \textbf{return} $\lambda_{opt} = \arg\min\limits_{q}L_q$.
		
	\end{algorithmic}
\label{alg:cv}
\end{algorithm}

\subsection{Feature screening}
\label{sec:fscreen}
When the number of features are large, there are two potentially undesirable consequences.  First, the computational time for the proposed estimation method is large, and second, there is a chance that some of the features are useless which may worsen the estimation quality. Therefore, we propose a feature screening procedure for removing some useless features before we apply the proposed graphon estimation method.

Let $\bm{D}=\{\tilde{d}_0(i,j)\}$ be the distance matrix based on the adjacency matrix $\bm{A}$, and $\bm{S} = \{\hat{s}(i,j)\}$ be the
dissimilarity matrix based on a single feature. We use the correlation between
$\{D_{ij}\}_{i\neq j}$ and $\{S_{ij}\}_{i\neq j}$ to describe the coherence of
features and graphon slices. If a feature is irrelevant, the correlation between $\bm{D}$ and $\bm{S}$ would be small. Since
$\{D_{ij}\}_{i\neq j}$ and $\{S_{ij}\}_{i\neq j}$ do not form a linear relation, we
use Kendall's $\tau$ correlation (or Spearman correlation). We will discuss the practical choice of threshold in Section~\ref{sec:experiment}.

\section{Theoretical Results}
\label{sec:theory}

In this section, we establish the asymptotic convergence of
the proposed estimator. For simplicity, we study the estimator without node features screening (Section \ref{sec:fscreen}). To accommodate important models such as stochastic block model,
we do not assume the graphon $w$ to be completely smooth.
Instead, we focus on the following family of functions.
\begin{definition}{3.1}(Bivariate piecewise Lipschitz graphon family)
	\label{def:bivariateLip}
	For any $\delta,L>0$, let $\mathcal{W}_{\delta;L}$ denote a family of piecewise Lipschitz graphon functions $w:[0,1]^2\rightarrow [0,1]$ such that (i) there exists $K\geq 1$ and $0=x_0<\ldots<x_K=1$ such that $\min\limits_{1\leq i\leq K}(x_{i}-x_{i-1})\geq\delta$; (ii) for any $(u_1,v_1)$ and $(u_2,v_2)\in [x_{i}-x_{i-1}]\times [x_{j}-x_{j-1}]$, $|w(u_1,v_1)-w(u_2,v_2)| \leq L(|u_1-u_2|+|v_1-v_2|)$.
\end{definition}
Similarly, our theory also allows piecewise Lipschitz form
for the feature function $f$ in the following sense.
\begin{definition}{3.2}(Piecewise Lipschitz feature family)
	\label{def:univariateLip}
	For any $\delta,L>0$, let $\mathcal{F}_{\delta;L}$ denote a family of piecewise Lipschitz functions $f:[0,1]\rightarrow \mathbb{R}^p$ such that (i) there exists $D\geq 1$ and $0=x_0<\ldots<x_D=1$ s.t. $\min\limits_{1\leq i\leq D} (x_{i}-x_{i-1})\geq\delta$; (ii) $\forall u, v\in [x_{i}-x_{i-1}]$, $|f_k(u)-f_k(v)| \leq L|u-v|, \ k=1,\ldots,p.$
\end{definition}

We further define the sub-Gaussian distribution.
\begin{definition}{3.2}(Sub-Gaussian distribution)
	\label{def:subgaussian}
	we say $X$ is sub-Gaussian($\sigma^2$) if $\E(X)=0$ and
	$\E[e^{sX}] \leq e^{\frac{\sigma^2 s^2}{2}} \text{ for } \forall s\in\mathbb{R}$.
\end{definition}
We have the following theorem
regarding the rate of convergence of the estimated probability matrix $\hat{\bm{P}}$.

\begin{theorem}{3.1} (Consistency of $\hat{\bm{P}}$)
	\label{thm:phat}
	Assume that (i) $w\in \mathcal{W}_{\delta_1;L^w}$ and $f\in \mathcal{F}_{\delta_2;L^f}$ with global constants $L^w$ and $L^f$; (ii) the length of the smallest common Lipschitz piece $\delta_{f\cap w}=\delta_{f\cap w}(n) := \min\limits_{i,j}\{ |I_i^f\cap I_j^w| : I_i^f\cap I_j^w\neq\phi \}$ satisfies $\lim\limits_{n\rightarrow\infty}\left(\delta_{f\cap w} \big/ \sqrt{\frac{\log n}{n}}\right) \rightarrow \infty$; (iii) $\E(e_{ik})=0$, and $e_{ik}\stackrel{iid}{\sim}$ sub-Gaussian($\sigma^2$) for all $i$ and $k$; (iv) $\|f(u)\|^2_2/p \leq M$ for any $u\in[0,1]$; (v) $h=C_0\sqrt{\frac{\log n}{n}}$ for any global constant $C_0$. Then for all $w\in \mathcal{W}_{\delta_1;L^w}$ and $\mathcal{F}_{\delta_2;L^f}$, we have
	\begin{equation*}
	\frac{1}{n^2} \| \hat{\bm{P}}-\bm{P} \|^2_F = \mathcal{O}_P\left( \sqrt{\frac{\log n}{n}} \right) + \lambda\mathcal{O}_P\left( \mathcal{M}(\sigma, p, n)\right),
	\end{equation*}
	where $\mathcal{M}(\sigma, p, n)$ is given as follows.
	\begin{itemize}
		\item[(a)] If $p>4\log n$,
		\begin{equation*}
		\mathcal{M}(\sigma, p, n) = \max\left\{ \sqrt{\frac{\log n}{n}}, \sigma \sqrt{\frac{\log n}{p}}, \sigma^2 \sqrt{\frac{\log n}{p}} \right\}.
		\end{equation*}
		\item[(b)] If $p\leq4\log n$,
		\begin{equation*}
		\mathcal{M}(\sigma, p, n) = \max\left\{ \sqrt{\frac{\log n}{n}}, \sigma \sqrt{\frac{\log n}{p}}, \sigma^2 \left(\frac{\log n}{p}\right) \right\}.
		\end{equation*}
	\end{itemize}
\end{theorem}

In our theorem, the best convergence rate of $\|\hat{\bm{P}}-\bm{P}\|_F^2/n^2$
is $\sqrt{\log n/n}$ which is sub-optimal when compared to
the minimax rate $\log n /n$.
However, as claimed by a recent work
\citep{zhang2015estimating},
$\sqrt{\log n /n}$ is the best rate obtained among existing non-combinatorial methods
for general graphon estimation without making strong assumptions such as
strict monotonicity of degree condition.
In our theorem, this rate can be achieved by both zero and nonzero $\lambda$ (with appropriate rate).
When $\lambda=0$, the node features play no role in the proposed estimation procedure.
This indicates that, in our theorem, we do not obtain any gain in terms of rate of convergence by the additional
usage of features. We hypothesize that this is largely due to the flexible modeling between features and hidden labels.
Besides, a similar conclusion is obtained by another recent work on community detection with node features \citep{zhang2015community},
which suggests that the real benefit of features is revealed in the finite sample performance.
As indicated evidently by our empirical study in Sections \ref{sec:experiment} and \ref{sec:application},
the usage of features (i.e., $\lambda>0$) improves the quality of estimation.
It is then of theoretical interest
to understand how large $\lambda$ could be accommodated by the proposed method without compromising the
overall rate of convergence.
Our theorem has shed light on this theoretical question.
The interplay of $\lambda, p, \sigma, n$ and their effects to the rate of convergence
can be easily seen by enforcing $\lambda\mathcal{M}(\sigma, p, n) = \mathcal{O}_P(\sqrt{\log n/n})$.
Here we give two examples under the setting of bounded $\sigma$.
When $p=\Omega(\log n)$ (high-dimensional setting), we require $\lambda$ to be $\mathcal{O}_P(\max(\sqrt{p/n},1))$.
As for $p=\mathcal{O}(\log n)$ (low-dimensional setting), we need $\lambda=\mathcal{O}_P(p/\sqrt{n\log n})$.

\section{Numerical Experiments}
\label{sec:experiment}

\subsection{Effects of node features}
\label{sec:example}
{\bf Setup:}
We generate a network $\bm{A}\in\{0,1\}^{n\times n}$ from a graphon $w(u,v)$ by the
two-step procedure described in (\ref{eqn:generate}). For each node, we
generate its features $\bm{X}_i\in\mathbb{R}^p$ from $f$ by (\ref{eqn:fpluse}).
If close $\bm{X}_i$'s imply close graphon slices, our method can leverage the information of features
and effectively improve the estimation.
One such assumption that guarantees this is that $f$ is smooth and monotonic.
As for when $f$ and the graphon has totally different structures, then inclusion of $s$ in the dissimilarity measure (i.e., choosing a large $\lambda$) will worsen the estimation.

The level of noise $\sigma$ also influences the effect of features. In order to compare the effects of different noise levels, we first standardize each feature $f_j$ by its standard deviation before adding noise. To be more specific,
\begin{equation*}
X_{ij} = f_j(u_i)/\mbox{sd}(f_j(U)) + e_{ij}, \ i=1,\ldots,n, \ j=1,\ldots,p,
\end{equation*}
where $U\sim\text{Uniform}(0,1)$ and $e_{ij}\stackrel{iid}{\sim}N(0,\sigma^2)$.

Finally, we define MSE and MAE to measure the performance of the estimation:
\begin{equation*}
\aligned
\text{MSE} &= \frac{1}{n^2}\sum_{i,j}(\hat{P}_{ij}-P_{ij})^2
\quad\mbox{and}\quad
\text{MAE} &= \frac{1}{n^2}\sum_{i,j}|\hat{P}_{ij}-P_{ij}|.
\endaligned
\end{equation*}

\noindent
{\bf An illustrative example:}
We will first use an illustrative example to show the effects of different node features. 

Consider the case of a single feature $f\in\mathbb{R}$ and a simple monotonic graphon $w(u,v)=(u+v)/2$ that is shown in Figure~\ref{fig:g0}. Any smooth and monotonic $f$ will be useful, for instance, $f = f_1(u)=\cos(\pi u)$, which approximates the true labels well. 
However, if $f = f_2(u)=\cos(2\pi u)$ or even $f=f_3(u)=\cos(4\pi u)$ (which are
non-monotonic and periodic), then close $\bm{X}_i$ and $\bm{X}_j$ may not well describe the
similarity of the corresponding graphon slices. Thus, $f_2$ and $f_3$ may not be as helpful.
Nevertheless, it turned out that using $f_2$ or $f_3$ can still help
locally and improve estimation when ${\lambda}$ is small since
they are smooth. More details are shown in
Figure~\ref{fig:g0effect}.

\begin{figure}[ht]
	\centering
	\subfloat{\includegraphics[height = 10cm, width = 10cm]{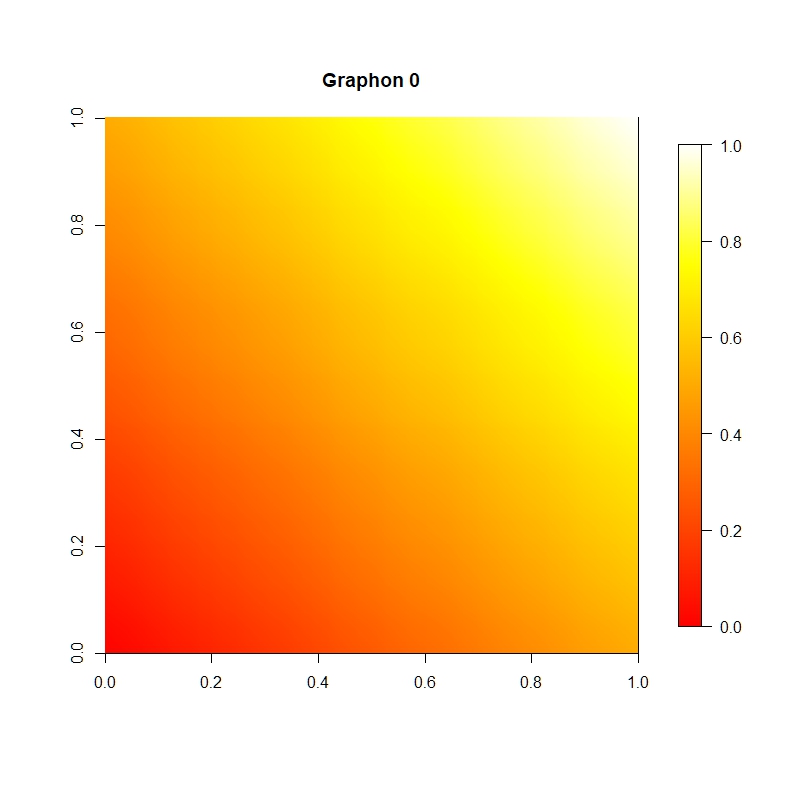}}
    \vspace*{-1.2cm}
	\caption{$w(u,v)=(u+v)/2$.}
	\label{fig:g0}
\end{figure}

\begin{figure}[ht]
	\centering
	\subfloat{\includegraphics[width = 0.5\textwidth]{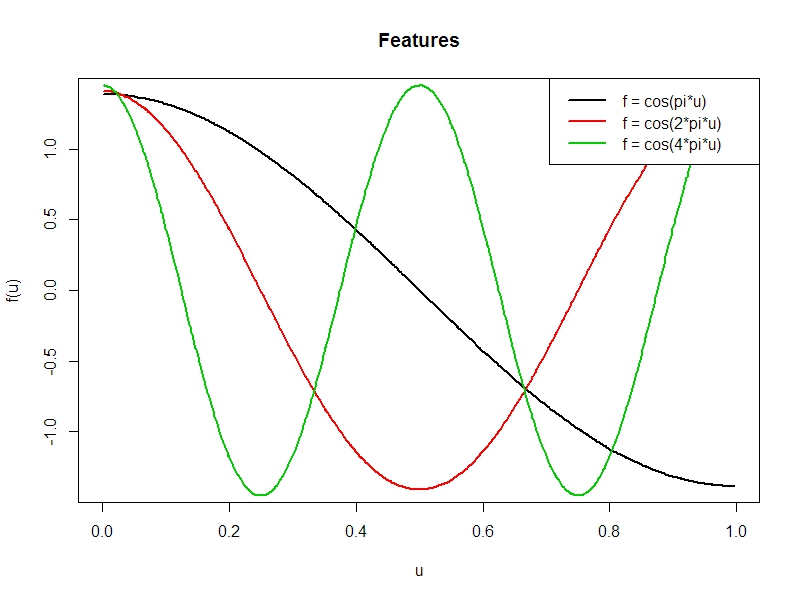}}
	\subfloat{\includegraphics[width = 0.5\textwidth]{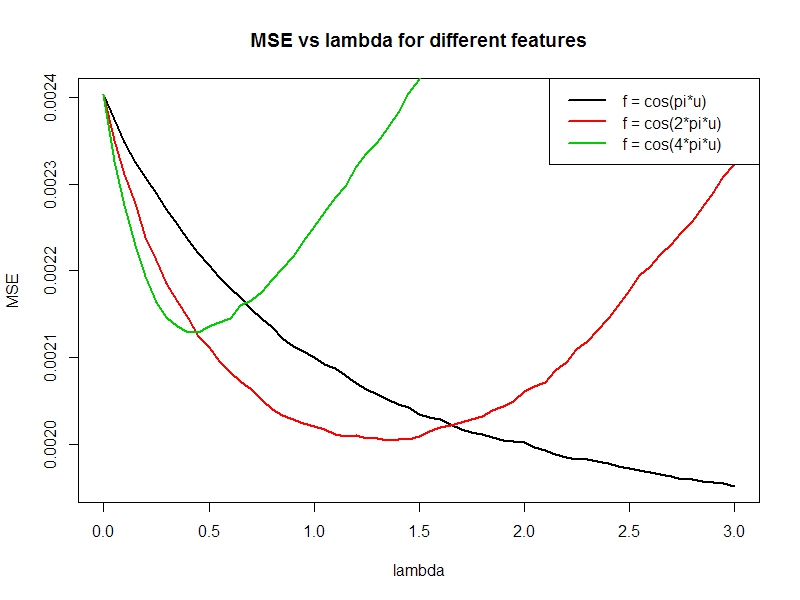}}
	\caption{The effect of different features $f_i(u)=\cos(2^{i-1}\pi u)$, $i=1,2,3$. Legends correspond to $f_1, f_2$ and $f_3$ from top to bottom. }
	\label{fig:g0effect}
\end{figure}

\noindent{\bf General graphons:}
Now, we study the effect of node features in the more general cases. We consider networks generated from the following four graphons that are used in the literature; see Figure~\ref{fig:generalgraphons} and Table~\ref{table:generalGraphon}. Graphon 1 is a stochastic block model (SBM) with $\lfloor\log n\rfloor$ blocks. Graphon 2 is periodic and low-rank. Graphons 3 and 4 are more general and both full-rank. Here the rank of a graphon is evaluated numerically on $\bm{P}$.
 
\begin{table}[t!]
\caption{Four general graphons used in numerical experiments.}
\label{table:generalGraphon}
\vspace*{-0.5cm}
	\begin{center}
			\begin{tabular}{cccc}
				\hline
				& Graphon $w(u,v)$ & Rank & local structure\\
				\hline
				$g_1$ & $k/(K+1)$ if $u,v\in (\frac{k-1}{K},\frac{k}{K})$; & $\lfloor\log n \rfloor$ & No\\
				& $0.3/(K+1) \text{ otherwise}.\ K=\lfloor\log n \rfloor$  & & \\
				\hline
				$g_2$ & $\frac{1}{2}\sin(5\pi(u+v-1)+1) + 0.5$ & 3 & No \\
				\hline
				$g_3$ & $1-\left[ 1+\exp\left\{ 15(0.8|u-v|)^{4/5} -0.1 \right\} \right]^{-1}$ & full & No\\
				\hline
				$g_4$ & $\frac{1}{3}(u^2+v^2)\cos(1/(u^2+v^2)) + 0.15$ & full & Yes \\
				\hline
			\end{tabular}
	\end{center}

\end{table}

\begin{figure}[ht]
	\centering
	\includegraphics[height = 8cm, width = 8cm]{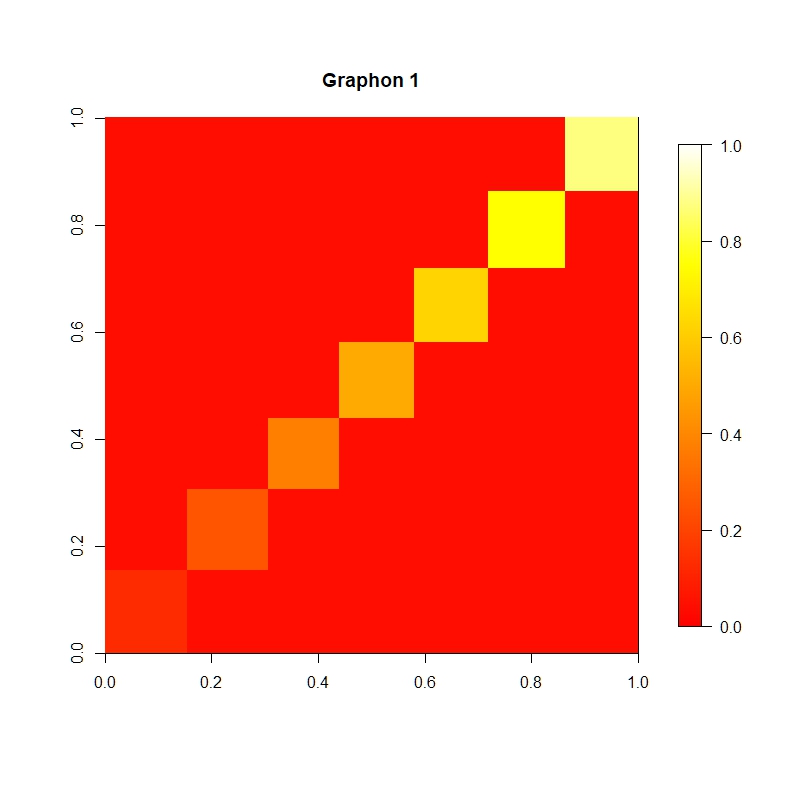}
	\includegraphics[height = 8cm, width = 8cm]{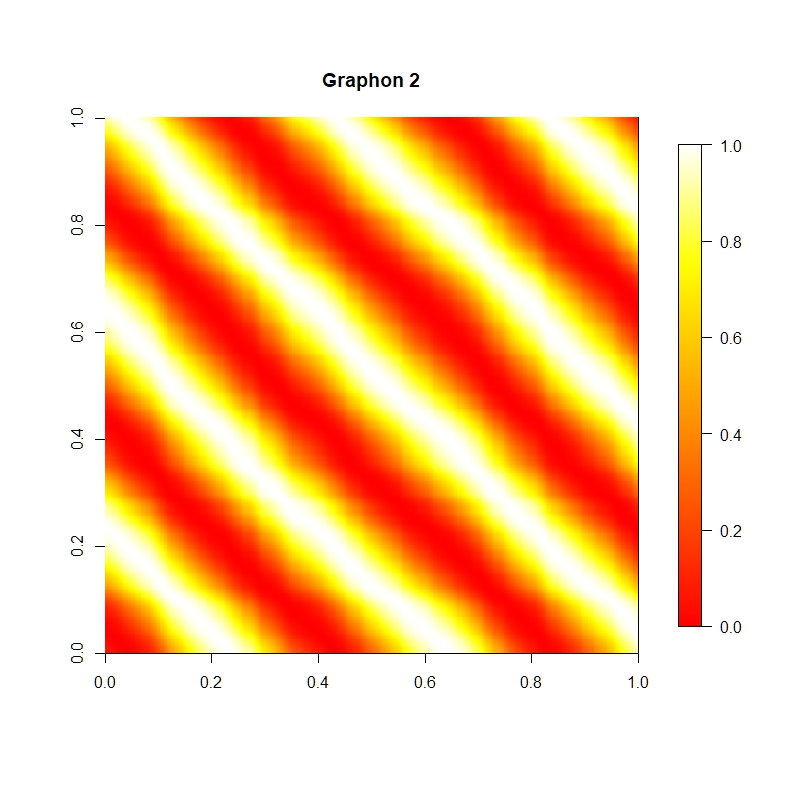}\\
    \vspace*{-2cm}
	\includegraphics[height = 8cm, width = 8cm]{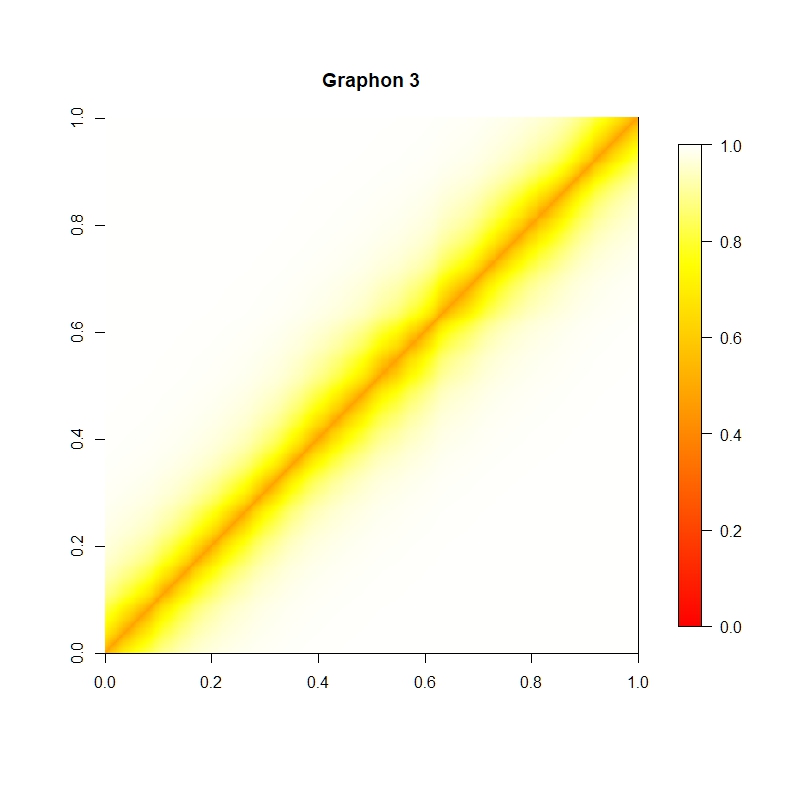}
	\includegraphics[height = 8cm, width = 8cm]{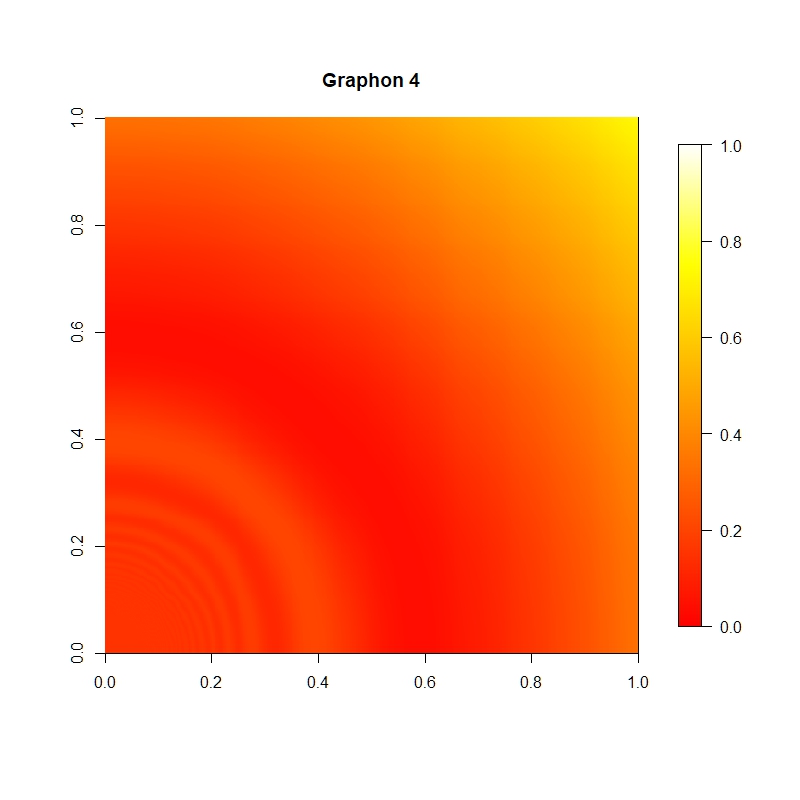}
    \vspace*{-1cm}
	\caption{Graphon visualizations when $n=2000$. From top-left to bottom-right (by row): $g_1$ to $g_4$.}
	\label{fig:generalgraphons}
\end{figure}

As for the node features, we select two non-monotonic functions and two monotonic ones:
\begin{equation*}
\aligned
f_1(u) &= \cos(2\pi(1-u)^2),  \ \ f_2(u) = 10u^2-12u+5, \\
f_3(u) &= \cos(\pi u),  \hspace{0.63in} f_4(u) = \Phi^{-1}(u),
\endaligned
\end{equation*}
and $f = (f_1,f_2,f_3,f_4)^T$, where $\Phi$ is the CDF function of standard normal distribution.

Figure~\ref{fig:generaleffects} shows the comparison of curves of MSE against ${\lambda}$ with different feature noise levels $\sigma=0, 0.1, \ldots, 0.5$ for $g_1$ to $g_4$, and each curve is an average from 20 repeated experiments.

\begin{figure}[ht]
	\centering\includegraphics[height=6cm, width=8cm]{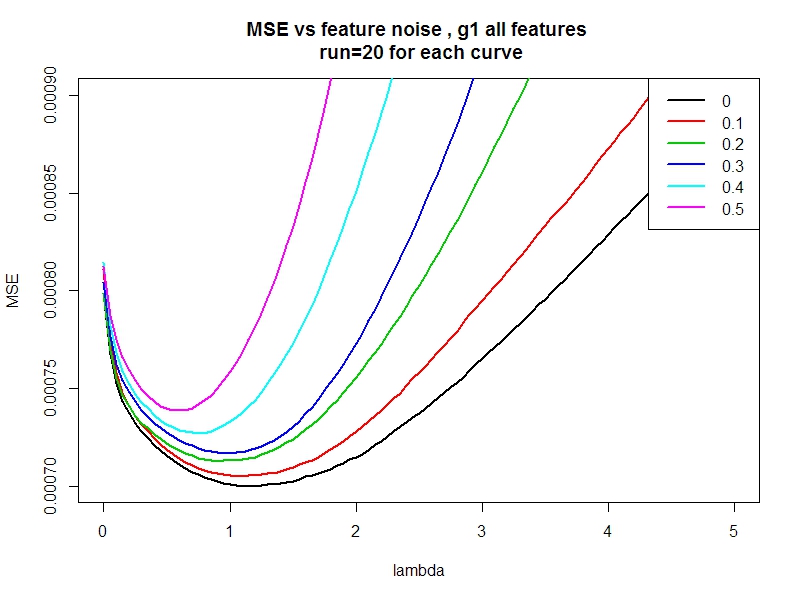}
	\centering\includegraphics[height=6cm, width=8cm]{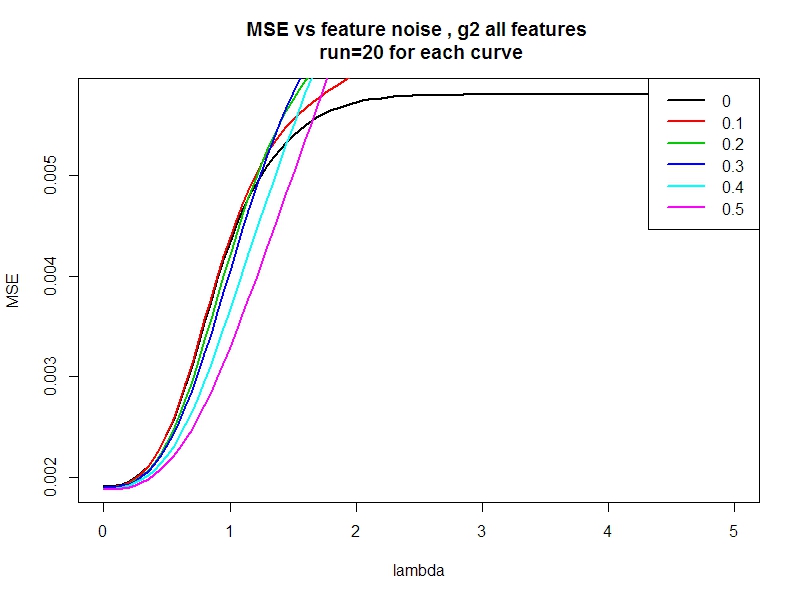}\\
	\centering\includegraphics[height=6cm, width=8cm]{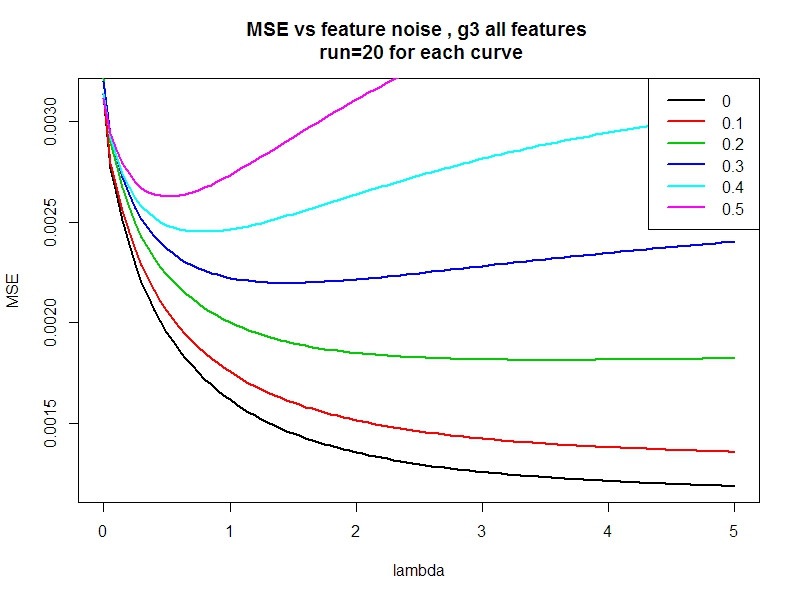}
	\centering\includegraphics[height=6cm, width=8cm]{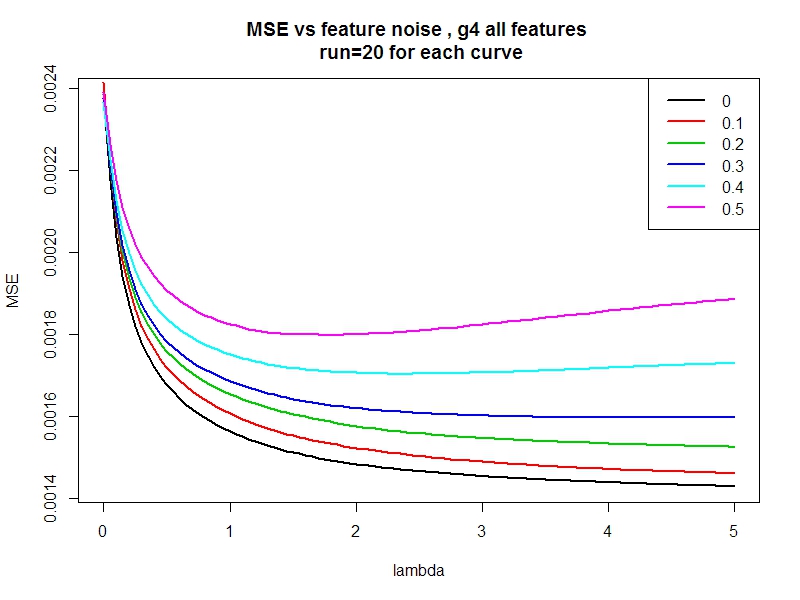}\\
	\caption{Effect of ${\lambda}$ with 20 independent experiments for each curve. From top-left to bottom-right (by row): $g_1$ to $g_4$. The starting point of each curve is ${\lambda}=0$, which is equivalent to the case without using features. Legends (from top to bottom) correspond to $\sigma=0, 0.1, \ldots, 0.5$ respectively.}
	\label{fig:generaleffects}
\end{figure}

Since $g_1$ is block-structured, using an appropriate amount of information from $f$ should improve the results, as smooth features make the neighborhoods more concentrated except near the boundaries. However when ${\lambda}$ is large, it will
pull nodes on the boundaries into wrong neighborhoods,
thus MSE can deteriorate rapidly. For $g_2$, although the graphon is smooth, close $\bm{X}_i$'s do not well correspond to similar graphon slices due to the periodic structure of $g_2$. In this case, the adjacency matrix $\bm{A}$ itself carries such strong information that adding features does not help that much. However, the MSE is not influenced a lot when ${\lambda}$ is close to 0. For $g_3$ and $g_4$, the effects of node features is significant since close features correspond to close graphon slices. There is a 20-30\% improvement in MSE with $p$-value$=0$. In particular, smooth node features can help better capture the local structure in $g_4$; see Appendix~\ref{appendix:local} for further details.

As expected, the performance becomes worse in general when the noise level of features increases. In practice, neither the pattern of features nor the noise level is known, therefore the cross-validation method is proposed for selecting an appropriate ${\lambda}$.

\subsection{Threshold for feature screening}
Recall that the feature screening procedure developed in Section~\ref{sec:fscreen} requires the specification of a threshold. Our numerical experiments show that 0.03 would be a reasonable choice of threshold. We tested it under Graphons 1 to 4 with feature being Gaussian noise. We performed 1000 independent trials to calculate the proportion of successful screen-out. The results are summarized in Table \ref{table:screening}. The probability of false positive (i.e., keeping a useless feature) is well controlled (approximately under 0.05) when we set the threshold as 0.03. In practice, the proposed feature screening mechanism should be combined with field knowledge to perform feature selection.

\begin{table}
	\caption{Feature screening with Gaussian noise feature.}
	\label{table:screening}
    \vspace*{-0.5cm}
	\begin{center}
		\begin{tabular}{c c c c c}
			\hline
			Graphon & $g_1$ & $g_2$ & $g_3$ & $g_4$ \\
			\hline
			$\#\{\tau<0.05\}/1000$ & 97.0\% &  99.9\% &  96.6\% &  94.7\%\\
			\hline
		\end{tabular}
	\end{center}
\end{table}

\subsection{Comparison with existing methods}
\label{sec:simulationcompare}
In this subsection we compare the performance of the proposed method FANS with other methods found in the literature.  These methods include 
\begin{itemize}
\item SBA: Stochastic Blockmodel Approximation of \cite{airoldi2013stochastic}, 
\item SAS: Sorting-and-Smoothing of \cite{chan2014consistent},
\item USVT: Universal Singular Value Thresholding of \cite{chatterjee2015matrix}, and 
\item NBS: Neighborhood Smoothing of \cite{zhang2015estimating}.
\end{itemize}

Given a graphon and a size $n$, we randomly generated $100$ independent realizations. For each realization we applied the above five methods to obtain the corresponding estimated graphons.  Since the SBA algorithm requires at least two independent graphs, we followed the way the authors conducted simulations in their paper and generated two of $\frac{n}{2}\times\frac{n}{2}$ graphs to make the comparison fair. For the SAS algorithm, we set the bandwidth $h=\log n$ as suggested in their paper. For NBS and FANS, we set $C_0=1$. For FANS specifically, feature screening (Section~\ref{sec:fscreen}) was performed and ${\lambda}$ was automatically chosen by cross-validation (Section~\ref{sec:cv}) before every fitting. 

We calculated the MSE and MAE for each estimated graphon. The results for $n=200$ and $500$ are summarized in Table~\ref{table:comparison}, where the averages and the standard errors of the calculated MSEs and MAEs are reported. The number at the end of each row is the $p$-value when the FANS method is compared with NBS using a paired one-sided $t$-test.  From this table, one can see that, when node features were available, FANS gave the best results (but we note that sometimes NBS gave similarly best results).  This is not surprising, as FANS is the only method that was designed to incorporate node feature information for graphon estimation.  When there is no such node feature present, FANS is similar to NBS, which gave extremely favorable results when comparing with the remaining methods.

\begin{table}[t!]
	\caption{Comparison with existing methods. Reported are the means and the standard errors (in parentheses) of MSEs and MAEs, based on 100 independent trials.  Numbers in brackets are the $p$-values of paired $t$-tests comparing the mean values between NBS and FANS. The lowest values are listed in bold.}
	\label{table:comparison}
    \vspace*{-0.5cm}
	\begin{footnotesize}
		\begin{center}
			\begin{tabular}{c l l l l l}
				\hline
				\bm{$n=200$} & SBA & SAS & USVT & NBS & FANS ($\sigma=0.3$) \\ 
				\hline
				\bm{$g_1$} MSE (SE) & 0.0112 (0.0015) & 0.0269 (0.0049) & 0.1679 (0.0013) &  \textbf{0.0020 (2.1e-4)} &  \textbf{0.0017 (1.9e-4)} [0.058] \\
				\hspace{0.12in} MAE (SE) & 0.0682 (0.0058) & 0.1039 (0.0138) & 0.3865 (0.0034) &  \textbf{0.0309 (0.0018)} &  \textbf{0.0296 (0.0012)} [0.058] \\
				\hline
				\bm{$g_2$} MSE (SE) & 0.0295 (0.0023) & 0.0983 (0.0160) & 0.0641 (0.0017) & \textbf{0.0040 (2.9e-4)} & \textbf{0.0042 (0.0018)} [0.1236] \\
				\hspace{0.12in} MAE (SE) & 0.1220 (0.0053) & 0.2709 (0.0255) & 0.1768 (0.0025) & \textbf{0.0479 (0.0017)} & \textbf{0.0489 (0.0079)} [0.1004] \\
				\hline
				\bm{$g_3$} MSE (SE) & 0.0158 (0.0015) & 0.0144 (3.3e-4) & 0.0122 (0.0021) & 0.0067 (3.7e-4) & \textbf{0.0039 (1.9e-4)} [0] \\
				\hspace{0.12in} MAE (SE) & 0.0684 (0.0057) & 0.0855 (0.0016) & 0.0738 (0.0106) & 0.0484 (0.0015) & \textbf{0.0327 (0.0011)} [0] \\
				\hline
				\bm{$g_4$} MSE (SE) & 0.0172 (0.0023) & 0.0044 (3.0e-4) & 0.1015 (0.0055) & 0.0044 (2.5e-4) & \textbf{0.0034 (2.9e-4)} [0] \\
				\hspace{0.12in} MAE (SE) & 0.0978 (0.0073) & 0.0545 (0.0018) & 0.2920 (0.0111) & 0.0526 (0.0015) & \textbf{0.0455 (0.0019)} [0] \\
				\hline
			\end{tabular}
		\end{center}
	\end{footnotesize}

	\begin{footnotesize}
		\begin{center}
			\begin{tabular}{c l l l l l}
				\hline
				\bm{$n=500$} & SBA & SAS & USVT & NBS & FANS ($\sigma=0.3$) \\ 
				\hline
				\bm{$g_1$} MSE (SE) & 0.0297 (5.5e-4) & 0.0210 (0.0046) & 0.1791 (7.0e-4) & \textbf{8.1e-4 (4.9e-5)} & \textbf{7.8e-4 (4.4e-4)} [5e-31]\\
				\hspace{0.12in} MAE (SE) & 0.0245 (0.0034) & 0.0782 (0.0120) & 0.4024 (0.0018) & \textbf{0.0201 (4.0e-4)} & \textbf{0.0198 (3.5e-4)} [4e-28]\\
				\hline
				\bm{$g_2$} MSE (SE) & 0.0154 (0.0014) & 0.0907 (0.0142) & 0.0617 (7.4e-4) & \textbf{0.0019 (7.7e-5)} & \textbf{0.0019 (7.6e-5)} [0.1728]\\
				\hspace{0.12in} MAE (SE) & 0.0899 (0.0044) & 0.2562 (0.0257) & 0.1679 (0.0016) & \textbf{0.0321 (6.3e-4)} & \textbf{0.0321 (6.3e-4)} [0.1238]\\
				\hline
				\bm{$g_3$} MSE (SE) & 0.0081 (7.8e-4) & 0.0136 (2.8e-4) & 0.0049 (4.2e-4) & 0.0031 (1.3e-4) & \textbf{0.0023 (7.5e-5)} [0]\\
				\hspace{0.12in} MAE (SE) & 0.0453 (0.0038) & 0.0839 (0.0015) & 0.0408 (0.0020) & 0.0293 (7.5e-4) & \textbf{0.0240 (4.7e-4)} [0]\\
				\hline
				\bm{$g_4$} MSE (SE) & 0.0099 (0.0015) & 0.0029 (2.1e-4) & 0.1008 (0.0032) & 0.0024 (8.8e-5) & \textbf{0.0017 (8.4e-5)} [0]\\
				\hspace{0.12in} MAE (SE) & 0.0768 (0.0061) & 0.0450 (0.0018) & 0.2895 (0.0067) & 0.0384 (7.9e-4) & \textbf{0.0326 (7.8e-4)} [0]\\
				\hline
			\end{tabular}
		\end{center}
	\end{footnotesize}
\end{table}

\section{Application to Real Data}
\label{sec:application}

In this section, we apply the proposed FANS method to a real-life dataset and illustrates its usefulness via both
visualization and a leave-one-out link prediction problem.
This dataset is related to friendship network and was collected by the National Longitudinal Study of
Adolescent Health (the AddHealth study), which can be downloaded from {\tt http://moreno.ss.uci.edu/data.html}. In this study students were asked to list their friends that they recently chatted with. Note that the original data are directed graphs, but we consider two students as friends if one named the other.
The whole dataset consists of 81 sub-datasets, each
containing either one or two schools. We analyzed one of them (``comm10'') which
contains 587 students from a single school. The three covariates recorded are
{\tt gender}, {\tt race} and {\tt grade}. We treated {\tt grade} as an ordinal variable. As for {\tt gender} and
{\tt race}, we convert them to a 0/1 vector. The $j$-th coordinate is $1$ if and only if the student belongs to the $j$-th category.

\subsection{Visualization}
The observed adjacency matrix is displayed in the bottom-right plot in Figure~\ref{fig:realcompare}. The block structure of the six communities that correspond to the six grades from 7 (upper-right) to 12 (bottom-left) is apparent. Here we also include a recently proposed community detection algorithm that also makes use of node features, JCDC, proposed by \citet{zhang2015community}. We chose the number of communities $K=6$ for JCDC. We applied the five methods mentioned in Section~\ref{sec:simulationcompare} to fit the underlying graphon, and JCDC to estimate the communities. 
The comparison is visualized in Figure~\ref{fig:realcompare}, where the nodes were sorted by grade. For the SAS algorithm, in addition to nodes sorted by grade, we also present the estimated graphon with nodes sorted by the empirical node degrees (which is the direct output from their method). From neither one of the two SAS plots can we observe any patterns of interest. The USVT method fails since it omitted all the singular values.

We can observe the block structures recovered from NBS, JCDC and FANS methods, but the latter two provided a much clearer view. However, JCDC seems to miss the two small communities near the lower corner.  On the other hand, the proposed FANS with $\lambda=0.1$ clearly distinguished all the six communities that correspond to the different grades. With the assistance of the node features, FANS was able to capture the two subtle communities omitted by all other methods near the lower corner.  Even when $\lambda=0$, FANS still outperformed NBS since FANS resolves the tie-issue discussed in Section~\ref{sec:method}.

\begin{figure}[!b]
	\vskip -0.05in
	\centering\includegraphics[width = 1\columnwidth]{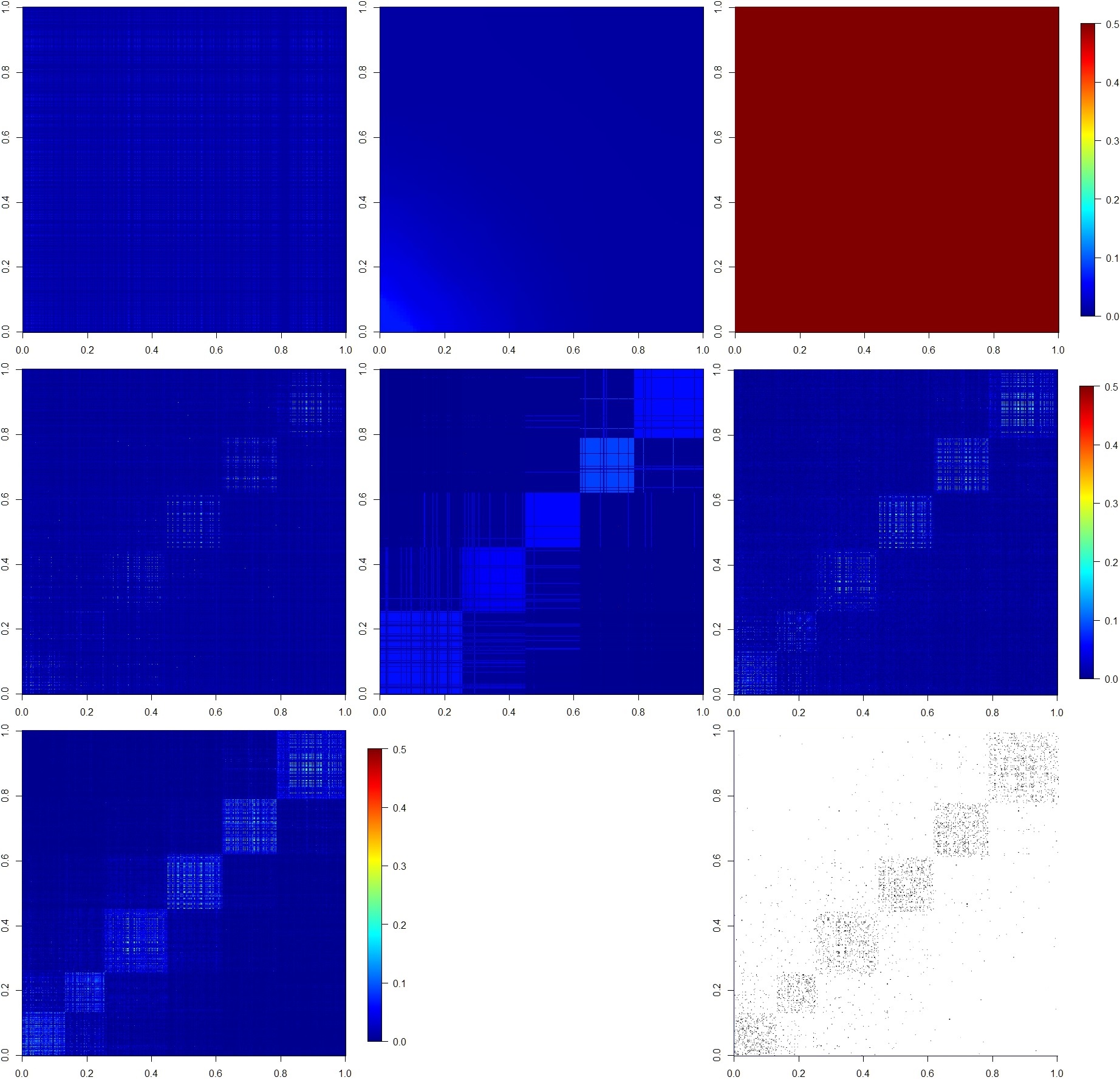}
	\caption{Estimated structures for the AddHealth ``comm10'' friendship network (first seven plots) and the observed adjacency matrix (bottom-right). From top to bottom (by row): SAS, SAS with nodes sorted by node degree, USVT, NBS, JCDC, FANS with $\lambda=0$, and FANS with $\lambda=0.1$.}
	\label{fig:realcompare}
	\vskip -0.2in
\end{figure}

We further sorted the nodes by gender. The estimated graphon by FANS is shown in Figure~\ref{fig:sortedgender}. The bottom-left block is the sub-network among females, and the upper-right corner forms the community among males. The other two parts are the cross-community connections. Within each block, there are 6 blocks that correspond to the six grades from 7 (upper-right) to 12 (bottom-left). One can see that the behaviors of male and female students are similar, and they are more likely to know each other in higher grades. This figure also illustrates that different graphons defined up to a measure preserving transformation correspond to the same network.

\begin{figure}[ht]
	\centering
	\centering\includegraphics[height=10cm, width=10cm]{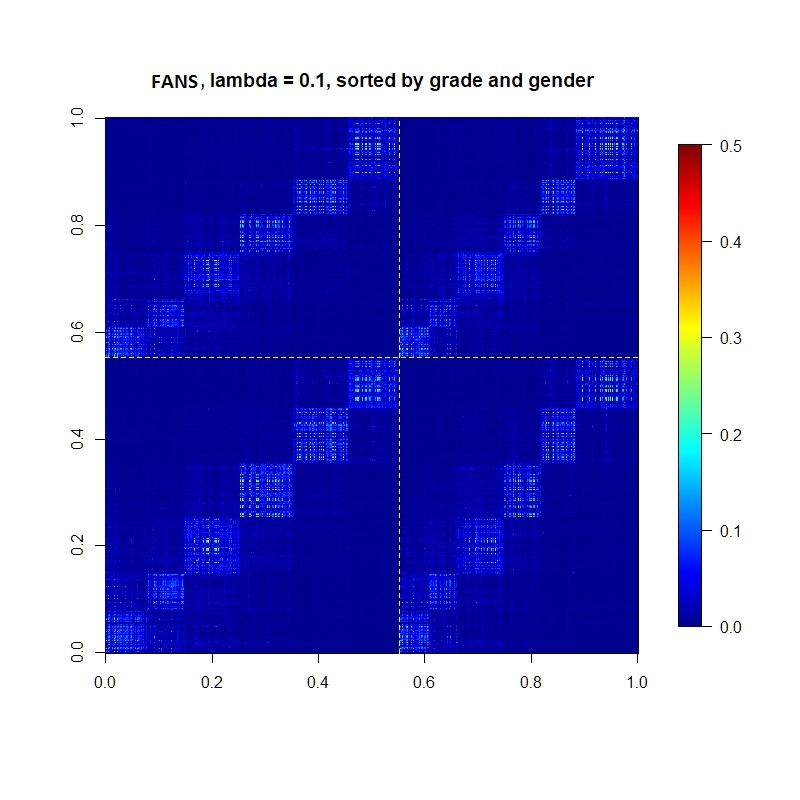}
    \vspace*{-1.2cm}
    \caption{Visualization of sub-networks in ``comm10''.}
	\label{fig:sortedgender}
\end{figure}

\subsection{Leave-one-out link prediction}
Since for this problem the true graphon is unknown, it is hard to quantitatively evaluate the quality of an estimated graphon. Therefore, we used a leave-one-out link prediction to compare FANS with NBS. In general, graphon estimation methods are not applicable to link prediction because they require a fully-observed network. However, FANS can be slightly modified such that the estimation of $\hat{P}_{ij}$ is independent of its observed value $A_{ij}$. This can be done by, in Step~1 of Algorithm~\ref{alg:proposed}, modifying $\langle \cdot, \cdot \rangle$ to $\langle \bm{A}_{i\cdot}, \bm{A}_{j\cdot} \rangle_{mod} := \sum_{k\neq i,j}A_{ik}A_{jk}$.  Then $\hat{P}_{ij}$ can be predicted completely independent of $A_{ij}$.

This is an element-wise procedure, and it can only predict one value at a time. Thus, we masked one pair of entries ($A_{ij}=A_{ji}$) each time and conducted a leave-one-out link prediction. 
We calculated the area under the receiver operating characteristic (ROC) curve to evaluate the link prediction. From Figure~\ref{fig:roc}, we can see that FANS with ${\lambda}=0.1$ performed the best.

\begin{figure}[ht]
	\centering
	\subfloat{\includegraphics[height=10cm, width=10cm]{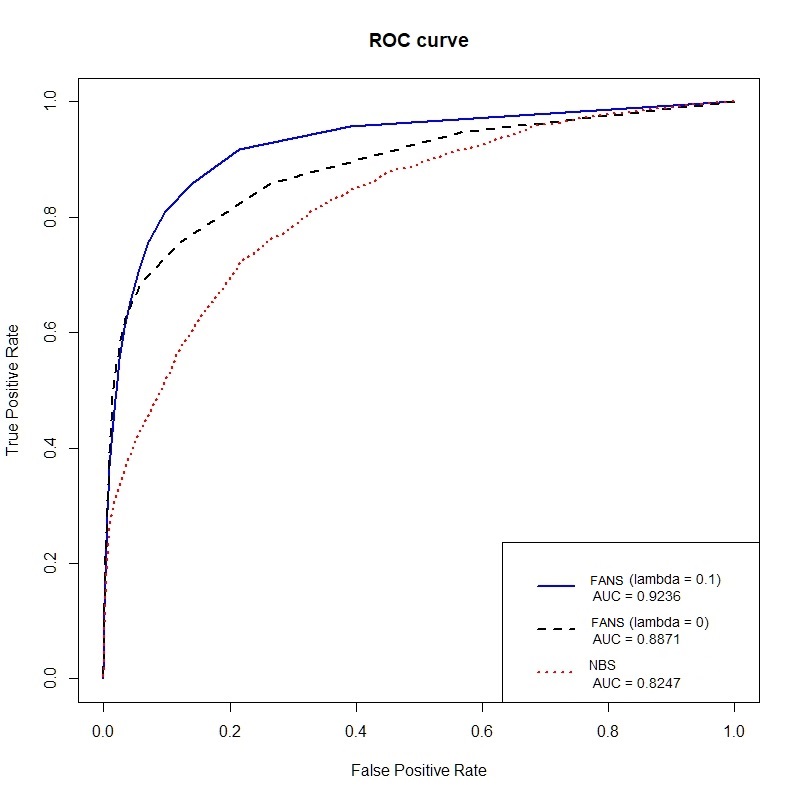}}
	\caption{ROC curves for NBS, FANS with $\lambda=0$, and FANS with $\lambda=0.1$.}
	\label{fig:roc}
\end{figure}

\section{Discussion}
\label{sec:discussion}
This paper developed a graphon estimation method that is capable of utilizing the information from both the observed adjacency matrix and node features. Under some mild regularity conditions, the consistency properties of the proposed method is established.  The rate of convergence is the same as without using node features, but in practice the proposed method can improve the estimation results in most cases.  Lastly, for a real world dataset the proposed method has benefits as it reflects more meaningful structures of a network and yields a higher link prediction accuracy.

\appendix

\section{Local Structure in Graphon 4}
\label{appendix:local}
\begin{figure}[ht]
	\centering\includegraphics[width = 1\columnwidth]{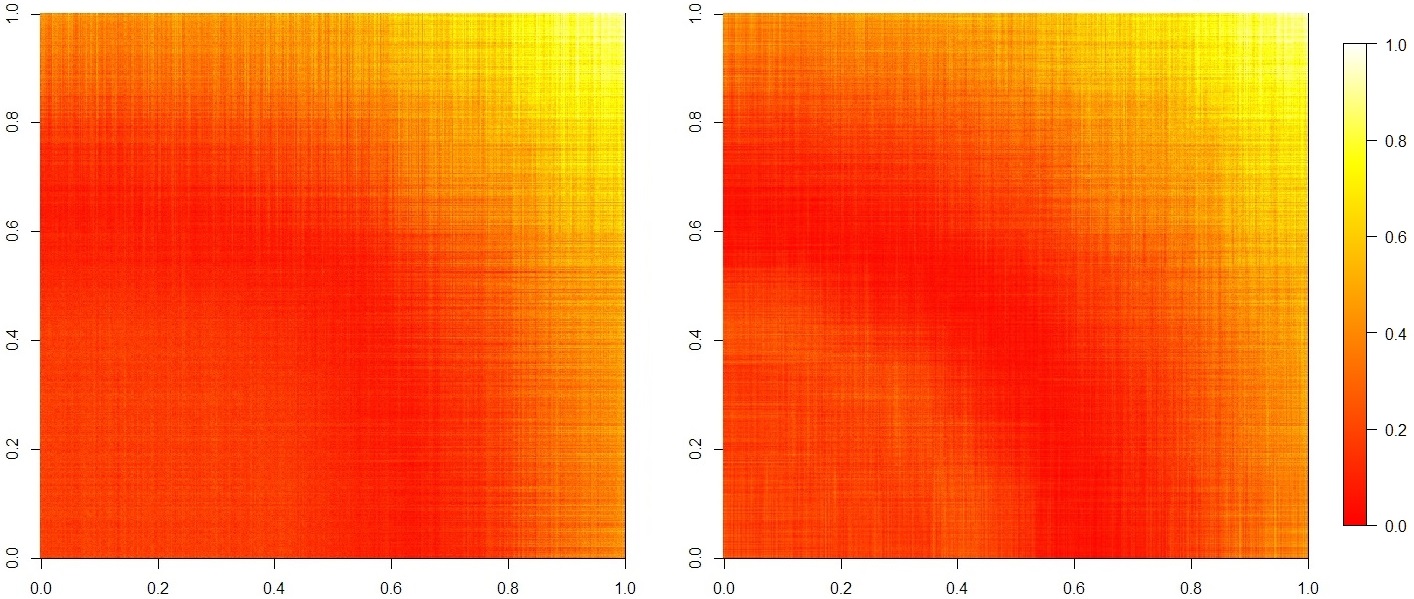}
	\caption{Local structure in $g_4$. Left: estimated graphon without using node features with the NBS method; right: estimated graphon with smooth node features obtained by the FANS method.}
	\label{fig:localstructure}
\end{figure}
This appendix provides an example to demonstrate the potential usefulness of incorporating node features into graphon estimation.  The bottom-left corner of $g_4$ presents a rich local structure that is difficult to be captured by the adjacency matrix. As shown in Figure~\ref{fig:localstructure}, the left estimation fails to detect any local structure since the signal carried by the adjacency matrix is relatively weak compared to the Bernoulli noise. On the other hand, the fitted graphon on the right of Figure \ref{fig:localstructure}, which utilizes smooth node features to assist the estimation, is able to capture such local information.

\section{Proof of Theorem 3.1}
This appendix presents technical arguments leading to the theoretical results in the paper.  In order to prove Theorem 3.1, if we write
\begin{equation*}
\hat{\bm{P}} = (\tilde{\bm{P}} +\tilde{\bm{P}}')/2 \ \text{ where} \ \tilde{\bm{P}} =  \frac{\sum_{i'\in N_i} A_{i'j}}{|N_i|},
\end{equation*}
then it suffices to prove the consistency of $\tilde{\bm{P}}$. In other words, we need to prove that
\begin{equation}
\label{eqn:Ptilde}
\frac{1}{n^2} \| \tilde{\bm{P}}-\bm{P} \|^2_F = \mathcal{O}_P\left( \sqrt{\frac{\log n}{n}} \right) + \lambda\mathcal{O}_P\left( \mathcal{M}(\sigma, p, n)\right).
\end{equation}
By triangle inequality for Frobenius norm, (\ref{eqn:Ptilde}) implies Theorem 3.1.

It is clear to observe that
\begin{equation*}
\frac{1}{n^2}\| \tilde{\bm{P}} - \bm{P} \|^2_F \leq \frac{1}{n}\max_{i}\Vert \tilde{\bm{P}}_{i\cdot} - \bm{P}_{i\cdot} \Vert_2^2,
\end{equation*}
so we only need to obtain a bound for the right-hand-side. Let's consider the following decomposition.
\begin{equation*}
\aligned
\frac{1}{n}\Vert \tilde{\bm{P}}_{i\cdot} - \bm{P}_{i\cdot} \Vert_2^2  &\leq \frac{1}{n}\sum_{j=1}^{n}\left\{ 2\left[ \frac{\sum_{i'\in N_i}(A_{i'j}-P_{i'j}) }{|N_i|}\right]^2 + 2\left[ \frac{\sum_{i'\in N_i}(P_{i'j}-P_{ij}) }{|N_i|}\right]^2 \right\} \\
&:=\frac{2}{n}\sum_{j=1}^{n}J_1(i,j) + \frac{2}{n}\sum_{j=1}^{n}J_2(i,j).
\endaligned
\end{equation*}

Let's first consider $J_1$. When $n>2$, we have
\begin{equation*}
\aligned
\frac{1}{n}\sum_{j=1}^{n}J_1(i,j)  &= \frac{1}{n|N_i|^2}\sum_{j=1}^{n} \left\{ \sum_{i'\in N_i}(A_{i'j}-P_{i'j})^2 + \sum_{i'\in N_i}\sum_{\substack{i''\in N_i\\i''\neq i'}}(A_{i'j}-P_{i'j})(A_{i''j}-P_{i''j}) \right\} \\
&\leq \frac{1}{|N_i|} + \frac{1}{n|N_i|^2} \sum_{j=1}^{n} \left\{ \sum_{\substack{i', i''\in N_i\\i'\neq i''}} (A_{i'j}-P_{i'j})(A_{i''j}-P_{i''j}) \right\}\\
&\leq \frac{1}{|N_i|} + \frac{1}{|N_i|^2} \sum_{\substack{i', i''\in N_i\\i'\neq i''}} \bigg| \frac{1}{n}\sum_{j=1}^{n} (A_{i'j}-P_{i'j})(A_{i''j}-P_{i''j}) \bigg|\\
&\leq \frac{1}{|N_i|} + \frac{1}{|N_i|^2} \sum_{\substack{i', i''\in N_i\\i'\neq i''}} \left[ \frac{1}{n-2}\bigg| \sum_{j\neq i',i''} (A_{i'j}-P_{i'j})(A_{i''j}-P_{i''j}) \bigg| +\frac{2}{n} \right]
\endaligned
\end{equation*}

For any $i'\neq i''$, by Bernstein's inequality, when $\varepsilon\in(0,1]$ and $n\geq 6$,
\begin{equation*}
\pr\left\{ \frac{1}{n-2}\bigg| \sum_{j\neq i',i''} (A_{i'j}-P_{i'j})(A_{i''j}-P_{i''j}) \bigg| \geq \varepsilon  \right\} \leq 2\exp\left\{ -\frac{\frac{1}{2}(n-2)\varepsilon^2}{1+\frac{1}{3}\varepsilon} \right\} \leq 2e^{-n\varepsilon^2/4}.
\end{equation*}
Taking a union bound over all $i'\neq i''$, we have
\begin{align*}
&\pr\left\{ \max\limits_{i; i'\neq i''\in N_i} \frac{1}{n-2}\bigg| \sum_{j\neq i',i''} (A_{i'j}-P_{i'j})(A_{i''j}-P_{i''j}) \bigg| \geq \varepsilon  \right\}\\
&\quad\leq
\pr\left\{ \max\limits_{i'\neq i''} \frac{1}{n-2}\bigg| \sum_{j\neq i',i''} (A_{i'j}-P_{i'j})(A_{i''j}-P_{i''j}) \bigg| \geq \varepsilon  \right\}
\leq 2n^2e^{-n\varepsilon^2/4}.
\end{align*}

Let $\varepsilon = \sqrt{(C_4+8)\frac{\log n}{n}}$ for some $C_4>0$ such that $\varepsilon\leq 1$, then we have
\begin{equation*}
\pr\left\{ \max\limits_{i; i',i''\in N_i} \frac{1}{n-2}\bigg| \sum_{j\neq i',i''} (A_{i'j}-P_{i'j})(A_{i''j}-P_{i''j}) \bigg| \geq \sqrt{(C_4+8)\frac{\log n}{n}} \right\} \leq 2n^{-C_4/4}.
\end{equation*}

Note that by definition of $N_i$, we have $|N_i|\geq C_0 (n-1)\sqrt{\frac{\log n}{n}} \geq C'_0\sqrt{n\log n}$ for some constants $C_0'>0$. Thus, when $n$ is large, with probability at least $1-2n^{-C_4/4}$, we have
\begin{equation*}
\aligned
\max\limits_{i} \frac{1}{n}\sum_{j=1}^{n}J_1(i,j)  &\leq \max\limits_{i}\left\{\frac{1}{|N_i|} + \frac{1}{|N_i|^2} \sum_{\substack{i', i''\in N_i\\i'\neq i''}} \left(  \sqrt{(C_4+8)\frac{\log n}{n}} +\frac{2}{n} \right)  \right\}\\
&\leq \frac{1}{C'_0\sqrt{n\log n}} + \sqrt{(C_4+8)\frac{\log n}{n}} +\frac{2}{n}\\
&= \tilde{C}_4\sqrt{\frac{\log n}{n}} 
\endaligned
\end{equation*}

We can see that the bound for $J_1$ is guaranteed by the size of $N_i$ as well as the nature of Bernoulli random variable. As for $J_2$, we have
\begin{equation*}
\aligned
\frac{1}{n}\sum_{j=1}^{n}J_2(i,j)  &=\frac{1}{n}\sum_{j=1}^{n}\left\{ \frac{\sum_{i'\in N_i}(P_{i'j}-P_{ij})}{|N_i|} \right\}^2\\
&\le\frac{1}{|N_i|}\sum_{i'\in N_i} \| \bm{P}_{i'\cdot}-\bm{P}_{i\cdot} \|^2/n,
\endaligned
\end{equation*}
due to Cauchy-Schwarz inequality. So, it suffices to bound $\max\limits_{i;i'\in N_i}\| \bm{P}_{i'\cdot}-\bm{P}_{i\cdot} \|^2/n$. In order to do so, we will need the following lemmas.

\begin{lemma}{1}
	Let $I_1^f,\ldots,I_D^f$ and $I_1^w,\ldots,I_K^w$ be the Lipschitz pieces for $f$ and $w$ respectively. Define a neighborhood (different from $N_i$) of label $u_i$ as 
	\begin{equation*}
	S_i(\Delta_n) = \{u_i\pm\Delta_n\} \cap I^f(u_i) \cap I^w(u_i)
	\end{equation*}
	where $I^f(u_i)$ is the Lipschitz piece of $f$ that contains $u_i$, $I^w(u_i)$ is the Lipschitz piece of $w$ that contains $u_i$, and $\Delta_n=(\tilde{C}_1+\sqrt{C_1+4})\sqrt{\frac{\log n}{n}}$ for any $C_1,\tilde{C}_1>0$ that satisfies $\Delta_n<\min\limits_{i,j}\{ |I_i^f\cap I_j^w| : I_i^f\cap I_j^w\neq\phi \}$ (so that $|S_i(\Delta_n)|>\Delta_n$) and $\sqrt{(C_1+4)\frac{\log n}{n}}\leq 1$. Then, we have
	\begin{equation*}
	\pr\left\{ \min\limits_{i}\frac{|\{\tilde{i}\neq i: u_{\tilde{i}}\in S(\Delta_n)\}| }{n-1} \geq \tilde{C}_1\sqrt{\frac{\log n}{n}} \right\} \geq 1-2n^{-C_1/4}
	\end{equation*}
\end{lemma}

Proof: By Bernstein's inequality, for any $\varepsilon\in(0,1]$ and $n\geq 6$, we have
\begin{equation*}
\pr\left\{ \bigg|\frac{1}{n-1} |\{\tilde{i}\neq i: u_{\tilde{i}}\in S(\Delta_n)\}| - |S_i(\Delta_n)| \bigg| \geq \varepsilon \right\} \leq 2\exp\left\{ -\frac{\frac{1}{2}(n-1)\varepsilon^2}{1+\frac{1}{3}\varepsilon} \right\} \leq 2e^{-n\varepsilon^2/4}.
\end{equation*}
Taking a union bound over all $i=1,\ldots,n$,
\begin{equation*}
\pr\left\{\max\limits_{i} \bigg|\frac{1}{n-1} |\{\tilde{i}\neq i: u_{\tilde{i}}\in S(\Delta_n)\}| - |S_i(\Delta_n)| \bigg| \geq \varepsilon \right\} \leq 2ne^{-n\varepsilon^2/4}.
\end{equation*}

Let $\varepsilon = \sqrt{(C_1+4)\frac{\log n}{n}}$ with any $C_1>0$ such that $\varepsilon\leq 1$, then we have
\begin{equation*}
\pr\left\{\max\limits_{i} \bigg|\frac{1}{n-1} |\{\tilde{i}\neq i: u_{\tilde{i}}\in S(\Delta_n)\}| - |S_i(\Delta_n)| \bigg| \geq \sqrt{(C_1+4)\frac{\log n}{n}} \right\} \leq 2n\cdot n^{-(C_1+4)/4} = 2n^{-C_1/4}.
\end{equation*}

Therefore, with probability at least $1-2n^{-C_1/4}$, we have
\begin{equation*}
\max\limits_{i} \bigg|\frac{1}{n-1} |\{\tilde{i}\neq i: u_{\tilde{i}}\in S(\Delta_n)\}| - |S_i(\Delta_n)| \bigg| \leq \sqrt{(C_1+4)\frac{\log n}{n}},
\end{equation*}
which implies that
\begin{equation*}
\aligned
\min\limits_{i} \frac{1}{n-1} |\{\tilde{i}\neq i: u_{\tilde{i}}\in S(\Delta_n)\}| &\geq \min\limits_{i}|S_i(\Delta_n)| - \sqrt{(C_1+4)\frac{\log n}{n}}\\
&\geq |\Delta_n| - \sqrt{(C_1+4)\frac{\log n}{n}}\\
&= \tilde{C}_1 \sqrt{\frac{\log n}{n}}
\endaligned
\end{equation*}

\begin{lemma}{2}
	Let parameter $h=C_0\sqrt{\frac{\log n}{n}}$ where $0<C_0\leq \tilde{C}_1$ with $\tilde{C}_1$, $\Delta_n$ and $S_i(\Delta_n)$ defined in Lemma 1. Recall that $\hat{d}^2(i,i') = \tilde{d}_0^2(i,i')+ \lambda \hat{s}^2(i,i')$ and $N_i := \{i': \hat{d}(i,i')\leq q_i(h) \}$. Then when $n$ is large enough, with high probability (tending to 1), we have
	\begin{equation*}
	\max_{i; i'\in N_i}\frac{1}{n}\| \bm{P}_{i'\cdot}-\bm{P}_{i\cdot} \|^2 \leq \tilde{C}\sqrt{\frac{\log n}{n}} + \lambda\mathcal{M}(\sigma, p, n)
	\end{equation*}
	where $\mathcal{M}(\sigma, p, n)$ is the bound of $\hat{s}^2(i,i')$ whose explicit form will be shown in Lemma 3.
	
\end{lemma}

Proof: First, we will show that for $i\neq j$, $\langle \bm{P}_{i\cdot},\bm{P}_{j\cdot}\rangle/n$ and $\langle \bm{A}_{i\cdot},\bm{A}_{j\cdot}\rangle/n$ are close. We will use $(\bm{P}^2/n)_{ij}$ and $(\bm{A}^2/n)_{ij}$ to represent the $(i,j)$-th entry of $\bm{P}^2/n$ and $\bm{A}^2/n$ respectively. Suppose $n>2$.

\begin{equation*}
\aligned
\big| (\bm{A}^2/n)_{ij} - (\bm{P}^2/n)_{ij} \big| &= \bigg| \frac{1}{n}\sum_{k=1}^{n} (A_{ik}A_{jk}-P_{ik}P_{jk}) \bigg|\\
&\leq \bigg| \frac{1}{n} \sum_{k\neq i,j} (A_{ik}A_{jk}-P_{ik}P_{jk}) \bigg| + \bigg| \frac{1}{n} \sum_{k= i,j} (A_{ik}A_{jk}-P_{ik}P_{jk}) \bigg| \\
&\leq \bigg| \frac{1}{n-2}\sum_{k\neq i,j} (A_{ik}A_{jk}-P_{ik}P_{jk}) \bigg| + \frac{4}{n}
\endaligned
\end{equation*}
By Bernstein's inequality, if $\varepsilon\in(0,1]$ and $n\geq 6$， we have
\begin{equation*}
\pr\left\{ \bigg| \frac{1}{n-2}\sum_{k\neq i,j} (A_{ik}A_{jk}-P_{ik}P_{jk}) \bigg| \geq \varepsilon \right\} \leq 2\exp\left\{ -\frac{\frac{1}{2}(n-2)\varepsilon^2}{1+\frac{1}{3}\varepsilon} \right\} \leq 2e^{-n\varepsilon^2/4}.
\end{equation*}
Taking a union bound over all $i\neq j$,
\begin{equation*}
\pr\left\{ \max\limits_{i\neq j} \bigg| \frac{1}{n-2}\sum_{k\neq i,j} (A_{ik}A_{jk}-P_{ik}P_{jk}) \bigg| \geq \varepsilon \right\} \leq 2n^2e^{-n\varepsilon^2/4}.
\end{equation*}

Let $\varepsilon = \sqrt{(C_2+8)\frac{\log n}{n}}$ for some $C_2>0$ such that $\varepsilon\leq1$, then we have
\begin{equation*}
\pr\left\{ \max\limits_{i\neq j} \bigg| \frac{1}{n-2}\sum_{k\neq i,j} (A_{ik}A_{jk}-P_{ik}P_{jk}) \bigg| \geq \sqrt{(C_2+8)\frac{\log n}{n}} \right\} \leq 2n^2\cdot n^{-(C_2+8)/4} = 2n^{-C_2/4}.
\end{equation*}

Therefore, with probability at least $1-2n^{-C_2/4}$, we have
\begin{equation*}
\max\limits_{i\neq j} \big| (\bm{A}^2/n)_{ij} - (\bm{P}^2/n)_{ij} \big| \leq \sqrt{(C_2+8)\frac{\log n}{n}}+\frac{4}{n} \leq 2\sqrt{(C_2+8)\frac{\log n}{n}} \text{ (when $n$ is large)}.
\end{equation*}

Next, we claim that for those $\tilde{i}$ such that $u_{\tilde{i}} \in S_i(\Delta_n)$, we have $\langle \bm{P}_{i\cdot},\bm{P}_{k\cdot}\rangle/n \approx \langle \bm{P}_{\tilde{i}\cdot},\bm{P}_{k\cdot}\rangle/n$. This is ensured by Lipschitz condition on $w$. In fact,
\begin{equation*}
\aligned
\max\limits_{i; \tilde{i}\in S_i(\Delta_n)} |(\bm{P}^2/n)_{ik}-(\bm{P}^2/n)_{\tilde{i}k}| &= \max\limits_{i; \tilde{i}\in S_i(\Delta_n)} |\langle \bm{P}_{i\cdot}- \bm{P}_{\tilde{i}\cdot},\bm{P}_{k\cdot}\rangle/n|\\
&\leq \max\limits_{i; \tilde{i}\in S_i(\Delta_n)} \|\bm{P}_{i\cdot}- \bm{P}_{\tilde{i}\cdot}\| \cdot \|\bm{P}_{k\cdot} \|/n\\
&\leq \max\limits_{i; \tilde{i}\in S_i(\Delta_n)} \sqrt{n(L^w|u_i-u_{\tilde{i}}|)^2} \cdot \sqrt{n} /n\\
&\leq L^w\Delta_n
\endaligned
\end{equation*}

Thus, combining the above two results, we can give a union bound for $\tilde{d}_0^2(i,\tilde{i})$ for all $i$ and any $\tilde{i}\in S_i(\Delta_n)$. With probability at least $1-2n^{-C_2/4}$, we have
\begin{equation*}
\aligned
\max\limits_{i; \tilde{i}\in S_i(\Delta_n)} \tilde{d}_0^2(i,\tilde{i}) &= \max\limits_{i; \tilde{i}\in S_i(\Delta_n)} \left\{ \max\limits_{k\neq i, \tilde{i}} |(\bm{A}^2/n)_{ik}-(\bm{A}^2/n)_{\tilde{i}k}| \right\}\\
&\leq \max\limits_{i; \tilde{i}\in S_i(\Delta_n)} \left\{ \max\limits_{k\neq i, \tilde{i}} |(\bm{P}^2/n)_{ik}-(\bm{P}^2/n)_{\tilde{i}k}| \right\} + 2\max\limits_{i\neq j} \big| (\bm{A}^2/n)_{ij} - (\bm{P}^2/n)_{ij} \big|\\
&\leq L^w\Delta_n + 4\sqrt{(C_2+8)\frac{\log n}{n}} = \tilde{C}_2 \sqrt{\frac{\log n}{n}}
\endaligned
\end{equation*}

We can also show that with probability at least $1-2n^{-C_3/4}-2n^{-C_3^*/4}$ ($C_3, C_3^*>0$),
\begin{equation}
\max\limits_{i; \tilde{i}\in S_i(\Delta_n)} \hat{s}^2(i,\tilde{i})\leq\mathcal{M}(\sigma, p, n),
\label{eqn:result1}
\end{equation}
where $\mathcal{M}(\sigma, p, n)$ is the bound that has the form derived in Lemma 3. Then with probability at least $1-2n^{-C_2/4}-2n^{-C_3/4}-2n^{-C_3^*/4}$,
\begin{equation}
\max\limits_{i; \tilde{i}\in S_i(\Delta_n)} \hat{d}^2(i,\tilde{i}) \leq \tilde{C}_2 \sqrt{\frac{\log n}{n}} + \lambda\mathcal{M}(\sigma, p, n).
\label{eqn:result2}
\end{equation}

Based on the definition of $N_i$ and that $h=C_0\sqrt{\frac{\log n}{n}}<\tilde{C}_1\sqrt{\frac{\log n}{n}}$, combining the results from Lemma 1 and (\ref{eqn:result2}), with probability at least $1-2n^{-C_1/4}-2n^{-C_2/4}-2n^{-C_3/4}-2n^{-C_3^*/4}$, we have
\begin{equation*}
\max\limits_{i; i'\in N_i} \hat{d}^2(i,i') \leq \tilde{C}_2 \sqrt{\frac{\log n}{n}} + \lambda\mathcal{M}(\sigma, p, n).
\end{equation*}

Finally, with probability $1-2n^{-C_1/4}-2n^{-C_2/4}-2n^{-C_3/4}-2n^{-C_3^*/4}$, for all $i, i'\in N_i$ uniformly,
\begin{equation*}
\aligned
\frac{1}{n}\| \bm{P}_{i'\cdot}-\bm{P}_{i\cdot} \|^2 &\leq |(\bm{P}^2/n)_{ii}-(\bm{P}^2/n)_{ii'}| + |(\bm{P}^2/n)_{i'i'}-(\bm{P}^2/n)_{i'i}| \\
&\leq |(\bm{P}^2/n)_{\tilde{i}i}-(\bm{P}^2/n)_{\tilde{i}i'}| + |(\bm{P}^2/n)_{\tilde{i}'i'}-(\bm{P}^2/n)_{\tilde{i}'i}| +4L^w\Delta_n\\
&\leq |(\bm{A}^2/n)_{\tilde{i}i}-(\bm{A}^2/n)_{\tilde{i}i'}| + |(\bm{A}^2/n)_{\tilde{i}'i'}-(\bm{A}^2/n)_{\tilde{i}'i}| + 4\tilde{C}_2\sqrt{\frac{\log n}{n}} +4L^w\Delta_n\\
&\leq 2\max\limits_{k\neq i,i'} |(\bm{A}^2/n)_{ik}-(\bm{A}^2/n)_{i'k}| + 4\tilde{C}_2\sqrt{\frac{\log n}{n}} +4L^w\Delta_n\\
&=2\tilde{d}_0^2(i,i') + 4\tilde{C}_2\sqrt{\frac{\log n}{n}} +4L^w\Delta_n\\
&\leq 2\hat{d}^2(i,i') + 4\tilde{C}_2\sqrt{\frac{\log n}{n}} +4L^w\Delta_n\\
&\leq \tilde{C} \sqrt{\frac{\log n}{n}} + \lambda\mathcal{M}(\sigma, p, n).
\endaligned
\end{equation*}
This completes the proof of Lemma 2.

\begin{lemma}{3}
	$\bm{X}_i=f(u_i)+e_i\in\mathbb{R}^p$ with $\E(e_{ik})=0$ and $e_{ik}\stackrel{iid}{\sim}$ sub-Gaussian($\sigma^2$). $f$ is piecewise Lipschitz with global constant $L^f$, and $\Vert f(u)\Vert_2^2/p\leq M$ for any $u\in[0,1]$ ($M$ is a global constant). Then with probability at least $1-2n^{-C_3/4}-2n^{-C_3^*/4}$ ($C_3, C_3^*>0$), we have
	\begin{itemize}
		\item[(i)] When $p>4\log n$,
		\begin{equation*}
		\max\limits_{i; \tilde{i}\in S_i(\Delta_n)} \hat{s}^2(i,\tilde{i})\leq \tilde{C}_3 \max\left\{ \sqrt{\frac{\log n}{n}}, \sigma \sqrt{\frac{\log n}{p}}, \sigma^2 \sqrt{\frac{\log n}{p}} \right\}
		\end{equation*}
		\item[(ii)] When $p\leq4\log n$,
		\begin{equation*}
		\max\limits_{i; \tilde{i}\in S_i(\Delta_n)} \hat{s}^2(i,\tilde{i})\leq \tilde{C}_3 \max\left\{ \sqrt{\frac{\log n}{n}}, \sigma \sqrt{\frac{\log n}{p}}, \sigma^2 \left(\frac{\log n}{p}\right) \right\}
		\end{equation*}
	\end{itemize}
\end{lemma}

Proof: Denote $\bm{\mu}_i = f(u_i)$. Similar to the proof of Lemma 2, we first need to show that $\langle \bm{X}_i, \bm{X}_j\rangle/p$ and $\langle \bm{\mu}_i, \bm{\mu}_j\rangle/p$ are close.
\begin{equation*}
|\langle \bm{X}_i,\bm{X}_j\rangle - \langle \bm{\mu}_i,\bm{\mu}_j\rangle| \leq |\langle \bm{\mu}_i,e_j \rangle| + |\langle\bm{\mu}_j, e_i \rangle| + |\langle e_i, e_j\rangle|
\end{equation*}

Since $e_{jk}\sim\text{sub-Gaussian}(\sigma^2)$, $\bm{\mu}_{ik}e_{jk}|u_i\sim\text{sub-Gaussian}(f_k(u_i)^2\sigma^2)$. By Hoeffding's inequality, for any $\varepsilon>0$, we have
\begin{equation*}
\pr\left\{ \bigg| \frac{1}{p} \langle \bm{\mu}_i,e_j \rangle \bigg| >\varepsilon \ \big| u_i \right\} \leq 2\exp\left\{ -\frac{p^2\varepsilon^2}{2\sigma^2\sum_{k=1}^{p}f_k(u_i)^2} \right\} \leq 2\exp\left\{ -\frac{p\varepsilon^2}{2\sigma^2 M } \right\}
\end{equation*}
where the RHS does not involve $u_i$. Therefore we have
\begin{equation*}
\pr\left\{ \bigg| \frac{1}{p} \langle \bm{\mu}_i,e_j \rangle \bigg| >\varepsilon \right\} \leq  2\exp\left\{ -\frac{p\varepsilon^2}{2\sigma^2 M } \right\}.
\end{equation*}
Taking a union bound over all $i$ and $j$ (it can be that $i=j$), and setting $\varepsilon = \sqrt{(C_3+4)\sigma^2M\frac{\log n}{p}}$ for any $C_3>0$, we have
\begin{equation*}
\pr\left\{ \max\limits_{i,j} \bigg| \frac{1}{p} \langle \bm{\mu}_i,e_j \rangle \bigg| >\sqrt{(C_3+4)\sigma^2M\frac{\log n}{p}} \right\} \leq 2n^2\exp\left\{ -\frac{(C_3+4)\sigma^2 M\log n}{2\sigma^2 M } \right\} = 2n^{-C_3/2}.
\end{equation*}

As for the $|\langle e_i, e_j\rangle|/p$ term, by Lemma 4, we can show that for $i\neq j$, $e_{ik}e_{jk}\sim$ sub-Exponential distribution with parameters $\nu=4\sigma^2$ and $\alpha = 4\sigma^2$.

[Case I:] When $0<\varepsilon\leq \nu = 4\sigma^2$, by Bernstein's inequality, we have
\begin{equation*}
\pr\left\{ \bigg| \frac{1}{p} \langle e_i,e_j \rangle \bigg| >\varepsilon \right\} \leq 2e^{-\frac{p\varepsilon^2}{2\nu^2}} = 2e^{-\frac{p\varepsilon^2}{32\sigma^4}}.
\end{equation*}

Taking a union bound over all $i\neq j$, and setting $\varepsilon =\sqrt{(C_3'+64)\sigma^4\frac{\log n}{p}}$, we have
\begin{equation*}
\pr\left\{ \max\limits_{i,j} \bigg| \frac{1}{p} \langle e_i,e_j \rangle \bigg| > \sqrt{(C_3'+64)\sigma^4\frac{\log n}{p}} \right\} \leq 2n^{-C_3'/32}.
\end{equation*}
The above inequality holds if $p>4\log n$ and $C_3'\in(0, \frac{16p}{\log n} - 64]$. Therefore, with probability at least $1-2n^{-C_3/4}-2n^{-C_3'/32}$, we have
\begin{equation*}
\max\limits_{i\neq j} |\langle \bm{X}_i,\bm{X}_j\rangle - \langle \bm{\mu}_i,\bm{\mu}_j\rangle| /p \leq 2\sqrt{(C_3+4)\sigma^2M\frac{\log n}{p}} + \sqrt{(C_3'+64)\sigma^4\frac{\log n}{p}}.
\end{equation*}

[Case II:] If $\varepsilon > \nu = 4\sigma^2$, Bernstein's inequality gives
\begin{equation*}
\pr\left\{ \bigg| \frac{1}{p} \langle e_i,e_j \rangle \bigg| >\varepsilon \right\} \leq 2e^{-\frac{p\varepsilon}{2\nu}} = 2e^{-\frac{p\varepsilon}{8\sigma^2}}.
\end{equation*}

Taking a union bound over all $i\neq j$, and setting $\varepsilon = (C_3''+16) \sigma^2 \frac{\log n}{p}$, we have
\begin{equation*}
\pr\left\{ \max\limits_{i,j} \bigg| \frac{1}{p} \langle e_i,e_j \rangle \bigg| > (C_3''+16) \sigma^2 \frac{\log n}{p} \right\} \leq 2n^{-C_3''/8}.
\end{equation*}
The above inequality holds if $p\leq4\log n$ and for any $C_3''>0$. Therefore, with probability at least $1-2n^{-C_3/4}-2n^{-C_3''/8}$, we have
\begin{equation*}
\max\limits_{i\neq j} |\langle \bm{X}_i,\bm{X}_j\rangle - \langle \bm{\mu}_i,\bm{\mu}_j\rangle| /p \leq 2\sqrt{(C_3+4)\sigma^2M\frac{\log n}{p}} + (C_3''+16)\sigma^2\frac{\log n}{p}.
\end{equation*}

Second, we will show that for $\tilde{i}$ such that $u_{\tilde{i}} \in S_i(\Delta_n)$, we have $\langle \bm{\mu}_i,\bm{\mu}_k\rangle/p \approx \langle \bm{\mu}_{\tilde{i}},\bm{\mu}_k\rangle/p$. This is ensured by Lipschitz condition on $f$. In fact, conditional on $\{u_i\}$,
\begin{equation*}
\aligned
\max\limits_{i; \tilde{i}\in S_i(\Delta_n)} |\langle\bm{\mu}_i,\bm{\mu}_k\rangle/p- \langle\bm{\mu}_{\tilde{i}},\bm{\mu}_k\rangle/p| &= \max\limits_{i; \tilde{i}\in S_i(\Delta_n)} |\langle \bm{\mu}_i- \bm{\mu}_{\tilde{i}},\bm{\mu}_{k}\rangle/p| \\
&\leq \max\limits_{i; \tilde{i}\in S_i(\Delta_n)} \| \bm{\mu}_{i}- \bm{\mu}_{\tilde{i}}\| \cdot \|\bm{\mu}_{k} \|/p\\
&\leq \max\limits_{i; \tilde{i}\in S_i(\Delta_n)} \sqrt{p(L^f|u_i-u_{\tilde{i}}|)^2} \cdot \sqrt{pM} /p\\
&\leq \sqrt{M}L^f\Delta_n
\endaligned
\end{equation*}
where $\sqrt{M}L^f\Delta_n$ does not depend on $u_i$. Now, we are ready to bound $\hat{s}^2(i,\tilde{i})$.

Under case I, when $p>4\log n$ and $C_3'\in(0, \frac{16p}{\log n} - 64]$, with probability at least $1-2n^{-C_3/4}-2n^{-C_3'/32}$, we have
\begin{equation*}
\aligned
\max\limits_{i; \tilde{i}\in S_i(\Delta_n)} \hat{s}^2(i,\tilde{i}) &= \max\limits_{i; \tilde{i}\in S_i(\Delta_n)} \left\{ \max\limits_{k\neq i,i'} |\langle \bm{X}_i,\bm{X}_k\rangle - \langle \bm{X}_j, \bm{X}_k\rangle|/p \right\}\\
&\leq \max\limits_{i; \tilde{i}\in S_i(\Delta_n)} \left\{ \max\limits_{k\neq i, \tilde{i}} |\langle\bm{\mu}_i,\bm{\mu}_k\rangle- \langle\bm{\mu}_{\tilde{i}},\bm{\mu}_k\rangle|/p  \right\} + 2\max\limits_{i\neq j} |\langle \bm{X}_i,\bm{X}_j\rangle - \langle \bm{\mu}_i,\bm{\mu}_j\rangle| /p \\
&\leq \sqrt{M}L^f\Delta_n + 2\left( 2\sqrt{(C_3+4)\sigma^2M\frac{\log n}{p}} + \sqrt{(C_3'+64)\sigma^4\frac{\log n}{p}} \right)\\
&= \sqrt{M}L^f(\tilde{C}_1+\sqrt{C_1+4}) \sqrt{\frac{\log n}{n}} +  (4\sigma\sqrt{(C_3+4)M} + \sigma^2\sqrt{(C_3'+64)}) \sqrt{\frac{\log n}{p}}\\ 
&= \tilde{C}_{31} \sqrt{\frac{\log n}{n}} + \tilde{C}_{32}\sigma \sqrt{\frac{\log n}{p}} + \tilde{C}_{33}\sigma^2 \sqrt{\frac{\log n}{p}}\\
&= \tilde{C}_3 \max\left\{ \sqrt{\frac{\log n}{n}}, \sigma \sqrt{\frac{\log n}{p}}, \sigma^2 \sqrt{\frac{\log n}{p}} \right\}.
\endaligned
\end{equation*}

Under case II, when $p\leq 4\log n$ and $C_3''>0$, with probability at least $1-2n^{-C_3/4}-2n^{-C_3''/8}$, we have
\begin{equation*}
\aligned
\max\limits_{i; \tilde{i}\in S_i(\Delta_n)} \hat{s}^2(i,\tilde{i}) &= \max\limits_{i; \tilde{i}\in S_i(\Delta_n)} \left\{ \max\limits_{k\neq i,i'} |\langle \bm{X}_i,\bm{X}_k\rangle - \langle \bm{X}_j, \bm{X}_k\rangle|/p \right\}\\
&\leq \max\limits_{i; \tilde{i}\in S_i(\Delta_n)} \left\{ \max\limits_{k\neq i, \tilde{i}} |\langle\bm{\mu}_i,\bm{\mu}_k\rangle- \langle\bm{\mu}_{\tilde{i}},\bm{\mu}_k\rangle|/p  \right\} + 2\max\limits_{i\neq j} |\langle \bm{X}_i,\bm{X}_j\rangle - \langle \bm{\mu}_i,\bm{\mu}_j\rangle| /p \\
&\leq \sqrt{M}L^f\Delta_n + 2\left( 2\sqrt{(C_3+4)\sigma^2M\frac{\log n}{p}} + (C_3''+16)\sigma^2\frac{\log n}{p} \right)\\
&= \sqrt{M}L^f(\tilde{C}_1+\sqrt{C_1+4}) \sqrt{\frac{\log n}{n}} +  4\sigma\sqrt{(C_3+4)M}\sqrt{\frac{\log n}{p}} + 2(C_3''+16)\sigma^2\frac{\log n}{p}\\
&= \tilde{C}_3 \max\left\{ \sqrt{\frac{\log n}{n}}, \sigma \sqrt{\frac{\log n}{p}},  \sigma^2 \left(\frac{\log n}{p}\right) \right\}.
\endaligned<
\end{equation*}

This completes the proof of Lemma 3.

\begin{lemma}{4}
	Let $Y_1,Y_2\stackrel{iid}{\sim}$sub-Gaussian($\sigma^2$) with $\E(Y_i)=0$. That is,
	\begin{equation*}
	\E[e^{sY_i}] \leq e^{\frac{\sigma^2 s^2}{2}}, \ \ \forall s\in\mathbb{R}.
	\end{equation*}
	Then, $Z = Y_1Y_2$ follows sub-Exponential distribution with $(\nu, \alpha) = (4\sigma^2, 4\sigma^2)$.
\end{lemma}

Proof:
\begin{equation*}
\aligned
\E[e^{sZ}] &= \E[e^{sY_1Y_2}] \\
&= 1 + \E(Y_1Y_2) + \sum_{k=2}^{\infty}\frac{s^k\E(Y_1Y_2)^k}{k!}\\
&= 1 + \sum_{k=2}^{\infty}\frac{s^k\E(Y_1^k)\E(Y_2^k)}{k!}\\
&\leq 1 + \sum_{k=2}^{\infty}\frac{s^k\E(|Y_1|^{2k})}{k!}\\
&\leq 1 + \sum_{k=2}^{\infty}\frac{s^k (2\sigma^2)^k k!}{k!}\\
&= 1 + (2\sigma^2 s)^2 \sum_{k=0}^{\infty}(2\sigma^2s)^k\\
&\leq 1 + 8\sigma^4s^2 \ \text{ (when $s\leq\frac{1}{4\sigma^2}$) }\\
&\leq e^{16\sigma^4s^2/2}.
\endaligned
\end{equation*}
Therefore, $Z=Y_1Y_2$ is sub-Exponential with $\alpha = 4\sigma^2$ and $\nu^2 = 16\sigma^4$.

\vspace{10mm}

Finally, we are ready to complete the main theorem. With probability at least $1-2n^{-C_1/4}-2n^{-C_2/4}-2n^{-C_3/4}-2n^{-C_3^*/4}-2n^{-C_4/4}$, we have
\begin{equation*}
\max\limits_{i}\frac{1}{n}\|\tilde{\bm{P}}_{i\cdot}-\bm{P}_{i\cdot}\|^2 \leq C\sqrt{\frac{\log n}{n}} + \lambda \mathcal{M}(\sigma, p, n),
\end{equation*}
which completes the proof of the main theorem.

\bibliography{graphon_reference}
\bibliographystyle{natbib}

\end{document}